\documentclass[aps,prl,reprint,superscriptaddress,longbibliography,floatfix]{revtex4-2} \usepackage[colorlinks=true,linkcolor=blue,citecolor=blue,urlcolor=blue]{hyperref}
\usepackage{microtype} % Improves typography
\microtypesetup{nopatch=footnote} %silences the footnote warning 
\usepackage{graphicx} % For graphics
\usepackage{amsfonts, amsthm, amsmath, amssymb, physics, upgreek, stackrel,bm} % Math environments and symbols
\usepackage{array, enumitem} % Tables and lists
\usepackage{verbatim, listings} % For comments and source code
\usepackage{dsfont, xr, lipsum} % Miscellaneous
\usepackage{placeins} % Contains \FloatBarrier command to control float placement
\usepackage[utf8]{inputenc}
\usepackage[T1]{fontenc}
\usepackage{lmodern}
\usepackage[english]{babel}
\usepackage{csquotes}
\usepackage{setspace}

\usepackage{pgfplots}
\usetikzlibrary{arrows.meta}
\pgfplotsset{compat=1.18}
\usepackage{bookmark}
\usepackage{tikz}

\usepackage[x11names,dvipsnames]{xcolor}
\usepackage{JanShortcuts}
\def\a{\ensuremath{{a}}} 
\def\b{\ensuremath{{b}}} 

\newtheorem{proposition}{Proposition}
\definecolor{StrangeGreen}{HTML}{58B359}
\usepackage{orcidlink}
\newcommand{\orcidLouw}{\orcidlink{0000-0002-5111-840X}}

\begin{document}
	
	\title{Lyapunov-controlled thermalization: an exact real-time example}
	\author{Jonas Loy}
    \affiliation{Faculty of Physics,
    Ludwig-Maximilians-Universit\"at M\"unchen,
    Munich, Germany}
	\author{Jan C. Louw \orcidLouw}
	\affiliation{Arbeitsgruppe Computing (AG C), Gesellschaft f\"ur wissenschaftliche Datenverarbeitung mbH G\"ottingen (GWDG), Burckhardtweg 4, D-37077 G\"ottingen, Germany}
	
	\begin{abstract}
		A verified Kubo-Martin-Schwinger (KMS) relation, after a drive has ceased, is not sufficient evidence to prove equilibrium. We demonstrate this in a large-$N$ large-$q$ Sachdev-Ye-Kitaev (SYK) quench protocol. Although all post-quench fermion correlators are exactly thermal, a second quench back to the initial Hamiltonian reveals hidden memory of the initial state. It is encoded in correlators connecting to times before the first quench. This memory is an extensive nonequilibrium (NEQ) witness with its decay rate being the Lyapunov exponent $\lambda_L$. Thus $\lambda_L$ sets the rate at which the state becomes effectively indistinguishable from a Gibbs state. Despite being a unitary interacting many-body system, the complete NEQ real-time evolution is obtained exactly in the large-$N$, large-$q$ limit, making the setup ideal for analytically studying thermalization.
	\end{abstract}
	\maketitle
    
	\emph{Introduction---.} Preparing a unitary quantum system out of equilibrium raises a basic question: when has it thermalized? Under unitary evolution, the von Neumann entropy is conserved, meaning the system cannot relax to a Gibbs state. 
	Instead, we define equilibrium operationally through accessible observables, typically few-body operators or sums thereof, as considered in the eigenstate thermalization hypothesis \cite{Mori:2017qhg, DAlessio03052016}. The KMS relation---an imaginary-time periodicity of the correlation functions---provides a rigorous notion of equilibrium \cite{Haag:1967sg}. As such, an established approach to studying thermalization is to ask when this periodicity is acquired \cite{Eberlein:2017jb,Louw2022Feb,Louw2026Sep,Osterkorn2026Sep, Bhattacharyya:2009uu, Keranen:2014lna, Ebrahim:2010ra}. 
	
	Studying in real time how a system becomes thermal is typically confined to special settings, such as integrable models \cite{Rigol:2016itf}, perturbative models \cite{Moeckel2008}, mean field theories \cite{aoki2014}, numerics for finite systems \cite{rigol2008thermalization} or effective descriptions of asymptotic regimes \cite{Liu:2018kfw}.
    Exact real-time access to interacting systems in the thermodynamic limit (large-$N$) throughout a NEQ protocol remains rare. The SYK model \cite{Kitaev2015,SachdevYe1992,Chowdhury2022Sep}, a strongly interacting system, provides such tractable dynamics \cite{Eberlein:2017jb,Louw2022Feb,Louw2026Sep,Osterkorn2026Sep,Perugu2025nov,Lensky:2020fqf}. For instance exact NEQ solutions, to the initial value problem, exist for $q$-body interacting SYK models to leading order in a $1/q$ expansion \cite{Louw2026Sep,Osterkorn2026Sep}. 
	
	Their KMS analysis shows that all fermionic correlation functions evaluated entirely after the quench immediately coincide with those of a thermal state. In this sense, the large-$q$ SYK model is stated to thermalize instantaneously. Neither of these papers, however, manages to study thermalization via the most intuitive measure: real-time thermalization of observables.
	
	In this letter, we fully solve the general multiquench problem exactly (shown in the SM). We focus however on the double quench problem which allows us to analytically study in real-time how a certain observable relaxes. The response to the second quench depends explicitly on the waiting time between the quenches and probes the state in terms of such previously inaccessible observables. Despite this exact thermal correlation sector, we find two unambiguous NEQ witnesses. The first is the expectation value of
	the initial SYK Hamiltonian which depends explicitly on the elapsed time between the two quenches. The second is an extensive thermodynamic entropy mismatch between the quenched state and its thermal reference state, which is necessary to avoid contradiction with the second law. This makes the NEQ witness macroscopically relevant.
	
	As such, the instantaneous thermalization of \cite{Eberlein:2017jb,Louw2022Feb,Louw2026Sep,Osterkorn2026Sep} is a statement about the post-quench correlator sector alone and not about the state. We find that the residual NEQ witness decays in real time. The rate is given by the Lyapunov exponent, typically related to scrambling, which thereby acquires an operational meaning as the rate of thermalization. 
	At low temperatures this rate approaches the Planckian bound $\sim 2\pi T$ \cite{Maldacena2016Aug}. Such Planckian scaling is also seen in black holes \cite{Lei2022Apr,Vishveshwara1970Aug,Carullo2021Apr,Louw2023Oct,Horiwitz2000Jun}, in numerical SYK studies \cite{Eberlein:2017jb,Osterkorn2026Sep,Bhattacharya2019Jul}, and argued via typicality \cite{Goldstein_2015}.
	
	Considering the connection to black holes above, our setup is naturally relevant to holographic approaches to thermalization.
	The IR regime of the SYK model admits a holographic description in terms of JT gravity \cite{Maldacena2016Nov, maldacenastanford2016conformal, Dhar:2018pii}. Importantly, our results are not obtained from the gravitational description but from the boundary real-time dynamics. As such, the model provides a particularly direct testbed for holographic thermalization \cite{Liu:2018holography, Arefeva:2012jp, Wu:2012rib}. A follow-up work extends the construction to several more general protocols while retaining analytic access \cite{jointSubmission}.

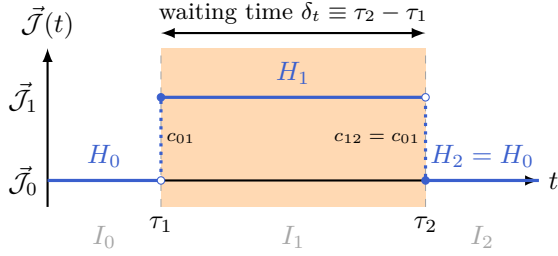
\begin{figure}[ht]
		\centering
		\begin{tikzpicture}[>=latex]
    \def\Ta{1.5}         % width of I_0
    \def\Tw{3.5}         % waiting time width
    \def\Tc{1.5}         % width of I_2
    \def\ya{0.0}         % level of J_0
    \def\yb{1.1}         % level of J_1
    % --- shaded intermediate interval ---
    \fill[orange!30] (0,-0.35) rectangle (\Tw,1.75);
    % --- axes ---
    \draw[->,thick] (-\Ta,0) -- (\Tw+\Tc,0) node[right] {$t$};
    \draw[->,thick] (-\Ta,-0.35) -- (-\Ta,1.75)
        node[above,align=center] {$\vec{\mathcal{J}}(t)$};
    % --- quench times ---
    \draw[dashed,gray!70] (0,-0.35) -- (0,1.75);
    \draw[dashed,gray!70] (\Tw,-0.35) -- (\Tw,1.75);
    \node[below] at (0,-0.35) {$\tau_1$};
    \node[below] at (\Tw,-0.35) {$\tau_2$};
    % --- the protocol:  H_0 ->  H_1 ->  H_2 =  H_0 ---
    \draw[very thick,RoyalBlue3] (-\Ta,\ya) -- (0,\ya);
    \draw[very thick,RoyalBlue3,dotted] (0,\ya) -- (0,\yb);
    \draw[very thick,RoyalBlue3] (0,\yb) -- (\Tw,\yb);
    \draw[very thick,RoyalBlue3,dotted] (\Tw,\yb) -- (\Tw,\ya);
    \draw[very thick,RoyalBlue3] (\Tw,\ya) -- (\Tw+\Tc,\ya);
    % open/closed circles at the jumps
    \filldraw[RoyalBlue3] (0,\yb) circle (1.4pt);
    \filldraw[RoyalBlue3] (\Tw,\ya) circle (1.4pt);
    \draw[RoyalBlue3,fill=white] (0,\ya) circle (1.4pt);
    \draw[RoyalBlue3,fill=white] (\Tw,\yb) circle (1.4pt);
    % --- level labels ---
    \node[left] at (-\Ta,\ya) {$\vec{\mathcal{J}}_{0}$};
    \node[left] at (-\Ta,\yb) {$\vec{\mathcal{J}}_{1}$};
    % --- Hamiltonian / interval labels ---
    \node[RoyalBlue3] at (-0.5*\Ta,\ya+0.32) {$ H_0$};
    \node[RoyalBlue3] at ( 0.5*\Tw,\yb+0.32) {$ H_1$};
    \node[RoyalBlue3] at ( \Tw+0.5*\Tc,\ya+0.32) {$ H_2= H_0$};
    \node[gray!70] at (-0.5*\Ta,-0.75) {$I_0$};
    \node[gray!70] at ( 0.5*\Tw,-0.75) {$I_1$};
    \node[gray!70] at ( \Tw+0.5*\Tc,-0.75) {$I_2$};
    % --- waiting time bracket ---
    \draw[<->,thick] (0,1.95) -- (\Tw,1.95)
        node[midway,above] {\small waiting time $\delta_t \equiv \tau_2-\tau_1$};
    % --- quench-strength angles ---
    \node[anchor=south west,inner sep=1pt] at (0.03,\ya+0.45)
        {\scriptsize $c_{01}$};
    \node[anchor=south east,inner sep=1pt] at (\Tw-0.03,\ya+0.45)
        {\scriptsize $c_{12} = c_{01}$};
\end{tikzpicture}
		\caption{Schematic of return-quench protocol. The time axis is partitioned into $I_0$, $I_1$ and $I_2$. 
			The quench strengths are encoded in the alignment of the coupling vectors, $c_{ab} \equiv \vec{\mathcal{J}}_a\cdot\vec{\mathcal{J}}_b/\mathcal{J}^2$. To simplify notation, we label functions by the intervals in which their time arguments lie, e.g. $\Jj(t_1)=\Jj_a$ or $c_{ab} = c(t_1,t_2)$ for $(t_1,t_2)\in I_a\! \times\! I_b$.
			%The double-quench protocol partitions the time axis into intervals: $I_0=(-\infty,0^-]$, $I_1=[0^+,\tau_2^-]$, and $I_2=[\tau_2^+,\infty)$.
		}
		\label{fig:protocol}
	\end{figure}

	\emph{Model \& protocol---.} To realize a nontrivial protocol, we consider a sum of mutually non-commuting SYK Hamiltonians $H^{(j)}$ with fixed, mutually uncorrelated disorder realizations
    \begin{equation}
		\label{the hamiltonian}
		H(t) \equiv H[\vec{\mathcal{J}}(t)]= \sum_{j=1}^{L} H^{(j)}[\mathcal{J}^{(j)}(t)].
	\end{equation}
	These are built from $N$ Majorana fermions $\psi_\mu$, $\mu=1,\dots,N$ with all-to-all independent random $q$-body interactions \cite{Maldacena2016Nov,jointSubmission} of effective coupling strength $\mathcal{J}^{(j)}(t)$. 
	Collected into a vector $\vec{\mathcal{J}}(t)$, with constant norm $\mathcal{J} = \vert\vec{\mathcal J}(t)\vert$ (see Sec.~\ref{app:multiquench-derivation}
    of the SM for the general multiquench protocol
    and solution), they encode the \emph{protocol}. We consider a return-quench protocol induced by piecewise constant couplings that are changed at $t=\tau_1,\tau_2$, separated by a waiting time $\delta_t \equiv \tau_2-\tau_1$, as sketched in Fig.~\ref{fig:protocol}.
    Since the pre and post quench Hamiltonians are non-commuting, they genuinely affect the state, rather than merely rescaling it.    

	Note that the observables we analytically have access to in SYK systems are typically the conserved quantities, such as the Hamiltonian, or operators built from the leading order in $1/N$ Green's functions; with some ingenuity one may also obtain next-to-leading-order corrections. Neither class serves as a witness here. The energy is conserved under evolution at constant coupling, hence $\expval{H_1}(t)$ is constant for $t\in I_1$ and tells us nothing about thermalization. Further, if the Green's functions satisfy a KMS relation, then so will any quantity built out of them. Fortunately, there is a way around this: the prequench Hamiltonian
	\begin{equation}
		\label{eq:probe}
		\expval{H_0}(\tau_2) \equiv \operatorname{Tr}[H_0\,\rho(\tau_2)] \xrightarrow[\tau_2\to\infty]{} \operatorname{Tr}[H_0 e^{-\beta_1 H_1}/Z].
	\end{equation}
	The above describes the real-time evolution of $\langle H_0\rangle$ not by varying $t$, but rather $\tau_2$. This provides a necessary criterion for equilibrium: were $\rho(\tau_2)$ thermal, the two would agree independent of the waiting time $\delta_t=\tau_2-\tau_1$. As such, any residual $\tau_2$ dependence of \eqref{eq:probe} is a direct NEQ witness. 
	
	We work in the thermodynamic limit $N\to\infty$, where the two-point functions $G(t_1,t_2)$ are determined by the Kadanoff-Baym equation (KBE), and higher-point functions can be constructed systematically from the two-point function \cite{Eberlein:2017jb, Gross:2017aos}.
	For Majorana fermions, the greater and lesser two-point functions obey $G^<(t_1,t_2)=G^>(t_1,t_2)^*$. 
	
	Writing $G^>(t_1,t_2)=-\frac{\imath}{2} e^{g(t_1,t_2)/q}$, with boundary conditions $g(t,t)=0$, we find our NEQ witness \eqref{eq:probe} via the Galitskii-Migdal relation \cite{Stefanucci2013Mar,Louw2022Feb}
	\begin{equation}
		\lim_{t_2 \to t} \partial_{t}g(t,t_2) = 2\imath \Tr [H(t)\rho(t)]q^2/N 
		\label{GMrelgen}
	\end{equation}
	evaluated at $t = \tau_2^+$.
	
	\emph{Dynamics---.} We consider the leading order in $1/q$, with the dynamics given by the KBE \cite{Louw2023Feb, Louw2022Feb, Eberlein:2017jb}
	\begin{align}
		\label{eq:first_order_g_main}
		\partial_{t_1} g(t_1,t_2) &= \imath\,\epsilon(t_1) + \int_{t_1}^{t_2} dt_3\, 2\vec{\mathcal{J}}(t_1)\cdot\vec{\mathcal{J}}(t_3) e^{g(t_1,t_3)}\\
		\epsilon(t)
		&\equiv \Im\!\left[\int_{-\infty}^{t} dt_3\, 2\vec{\mathcal{J}}(t)\cdot\vec{\mathcal{J}}(t_3) e^{g(t,t_3)}\right] \label{eq:closure_main}\\
		&=  2\Tr [H(t)\rho(t)]q^2/N. 
		\label{GMrel}
	\end{align}
	
	In the final line above we have made use of \eqref{GMrelgen} to relate $\epsilon(t)$ to the energy and thus interval-wise constant $\epsilon_a \propto \expval{H_a}(t)$. It provides the equal-time boundary condition and is determined self-consistently through \eqref{eq:closure_main}. Acting with $\partial_{t_2}$ on \eqref{eq:first_order_g_main} reveals a Liouville-type equation, which is solvable for constant couplings.

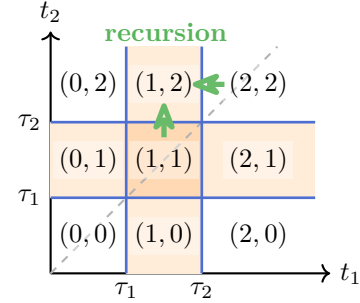
\begin{figure}[ht]
		\centering
		\begin{tikzpicture}[scale=1]
    \def\T{1}
    \def\R{2.5}   % right edge of the wide last column (in units of \T)
    % axes - along the bottom and left edges of the grid
    \draw[->, thick] (-1*\T,-1*\T) -- ({(\R+0.2)*\T},-1*\T) node[right] {$t_1$};
    \draw[->, thick] (-1*\T,-1*\T) -- (-1*\T,2.2*\T) node[above] {$t_2$};

    % orange cross highlight
    \fill[orange!15] (0*\T,1*\T) rectangle (1*\T,2*\T);   % (1,2)
    \fill[orange!30] (0*\T,0*\T) rectangle (1*\T,1*\T);   % (1,1)
    \fill[orange!15] (1*\T,0*\T) rectangle (\R*\T,1*\T);  % (2,1)
    \fill[orange!15] (-1*\T,0*\T) rectangle (0*\T,1*\T);  % (0,1)
    \fill[orange!15] (0*\T,-1*\T) rectangle (1*\T,0*\T);  % (1,0)
        % grid
    % highlight diagonal
    \draw[black!30,thick, dashed] (-1*\T,-1*\T) -- (2*\T,2*\T);
    \foreach \x in {0,1} {
        \draw[RoyalBlue3!90, line width=1pt] (\x*\T,-1*\T) -- (\x*\T,2*\T);
    }
    \foreach \y in {0,1} {
        \draw[RoyalBlue3!90, line width=1pt] (-1*\T,\y*\T) -- ({\R*\T},\y*\T);
    }
    % cell labels
    \foreach \a/\x in {0/-0.5,1/0.5,2/{0.5*(1+\R)}} {
            \foreach \b/\y in {0/-0.5,1/0.5,2/1.5} {
                \node[fill=white, fill opacity=0.4, text opacity=1, inner sep=2pt] 
                    at ({\x*\T},\y*\T) {$(\a,\b)$};
            }
        }
    % arrows into (1,2) - drawn last so labels cannot cover them
    \draw[line width=2pt, -{Stealth[length=14pt, open, round]} , StrangeGreen!90] (0.5*\T,0.8*\T) -- (0.5*\T,1.25*\T);
    \draw[line width=2pt, -{Stealth[length=14pt, open, round]} , StrangeGreen!90] (1.3*\T,1.5*\T) -- (0.88*\T,1.5*\T);
    % label for arrows
    \node[StrangeGreen] at ({0.15*(1+\R)*\T},2.2*\T) {\textbf{recursion}};
    % tick labels on the axes
    \node[below] at (0,-1*\T) {$\tau_1$};
    \node[below] at (\T,-1*\T) {$\tau_2$};
    \node[left] at (-1*\T,0) {$\tau_1$};
    \node[left] at (-1*\T,\T) {$\tau_2$};
\end{tikzpicture}
		\caption{Partition of the $(t_1,t_2)$-plane into blocks $I_a \times I_b$. The boundary conditions are imposed on the blue connecting lines. 
			The green arrows illustrate the construction of an off-diagonal block from its two neighboring blocks. The diagonal parameters $\theta_a$ are determined self-consistently through \eqref{eq:closure_main}. The integral therein depends on all data $\theta_{b<a}$. This further constraint determines any $\theta_a$ in terms of initial conditions $\theta_0$ self-consistently.}
		\label{fig:grid1}
	\end{figure}

	The $(t_1,t_2)$-plane is divided into blocks $I_a\! \times \!I_b$, as illustrated in Fig.~\ref{fig:grid1}. Inside each block, the couplings are constant, so the equation of motion reduces to a simpler constant-coupling problem.  
	
	The prequenched state is thermal at a temperature $T_0$ 
	\begin{equation}
		\label{eq:diagonal_solution_main}
		e^{g_{00}(t_1,t_2)}
		= \left[\frac{
			2\theta_0 T_0/\mathcal{J}
		}{\cosh\!\left[
			2\theta_0 T_0(t_1-t_2)
			+\imath\theta_0\right]}
		\right]^2,
	\end{equation}
	that is only dependent on time differences $g_{0}(t_1-t_2) \equiv g_{00}(t_1,t_2)$. Imposing the boundary conditions yields $2T_0\theta_0=\mathcal{J}\cos\theta_0$. Via \eqref{GMrelgen} we find the relationship to the initial energy $\epsilon_0 = -2\mathcal{J} \sin\theta_0$, and so $\sin\theta_0$ is a dimensionless energy measure. Infinite temperature corresponds to $\sin\theta_0=0$, while $\sin\theta_0=1$ corresponds to $T_0=0$. 
	
	Remarkably the diagonal solution $g_{aa}$ of \eqref{eq:first_order_g_main} continues to be of the same form as \eqref{eq:diagonal_solution_main} with the only difference being the replacement $(\theta_0,T_0) \to (\theta_a,T_a)$. In the thermodynamic limit, any ordinary SYK $n$-point function follows from the two-point function and thereby coincides with its thermal reference. In this way the system appears to be instantaneously thermal \cite{Eberlein:2017jb, Louw2022Feb}.
	
	The solution in the mixed-interval (off-diagonal) blocks can also be found by using the interval-wise constant $\theta_a$, continuity of $g$ across different intervals and the known solutions $g_{aa}$.
	Relating $\sin\theta_a$ to the integral in \eqref{eq:closure_main}, with the solution interval-wise inserted, determines $\theta_a$ in terms of $\theta_{b<a}$. 
	For any initial temperature $T_0$, this induces an algorithm for $g_{ab}$ as presented in Fig.~\ref{fig:grid1}.
	
	\emph{Results---.} Our thermalization witness \eqref{eq:probe} may now be simply calculated using the relation \eqref{GMrelgen} on $g_{aa}$ in interval $I_a$; explicitly via $\epsilon_a = -2\mathcal{J} \sin\theta_a$ and \eqref{GMrel} $\epsilon_a(t) = 2 \expval{H_a}(t) q^2/N$. 
	The response to the quench $H_0 \to H_1$ is
	\begin{equation}
		\label{single_quench_result}
		\epsilon_1 = c_{01}\, \epsilon_0 \, ,\quad c_{01}\equiv \vec{\mathcal{J}_0}\cdot \vec{\mathcal{J}_1}/\mathcal{J}^2
	\end{equation}
	as was shown in \cite{Louw2026Sep}. Here $c_{ab} \in [0,1]$
	measures the strength of the quench $H_a \to H_b$. In particular, the response \eqref{single_quench_result} is independent of the time of this first quench $\tau_1$.
	
	After the return quench $H_1 \to H_0$ ($c_{12} = c_{01}$), at $t_1=\tau_2^+$, \eqref{eq:closure_main} receives contributions from both the
	diagonal solution $g_{11}$ and the mixed-interval solution $g_{10}$, and therefore depends on both $\theta_1$ and $\theta_0$. 
	The mixed-interval solution is given by
	\begin{align}
		\label{recursive relation in exp form}
		e^{g_{01}(t_1, t_2)} =& \frac{e^{g_{0}(t_1-\tau_1)}e^{ g_{1}(\tau_1-t_2)}
		}{\left[1-c_{01}\, V_{0}(t_1)\, V_{1}^*(t_2)\right]^2},
	\end{align}
	where $V_{b}(t) \equiv [\dot{g}_{b}(0)-\dot{g}_{b}(t-\tau_1)]/(2\mathcal{J})$. 	
    Finally, having solved for the Green's function we obtain
	\begin{equation}
    \boxed{
		\epsilon_2(\tau_2)
		=c_{12}\,\epsilon_1+[\epsilon_0-c_{12}\,\epsilon_1]
		\vert e^{g_{11}(\tau_2,\tau_1)}\vert \propto \Tr[H_0 \rho(\tau_2)]
		\label{eq:exact_return_result}
        }
	\end{equation}
    via the integral \eqref{eq:closure_main} evaluated at $t=\tau_2^+$, see~\ref{SM-recursiveresolution}--\ref{SMss:self-con} for details. Here, varying $\tau_2$ probes the time evolution of $\Tr[H_0 \rho(t)]$, whether or not the second quench is performed.
	The second term in \eqref{eq:exact_return_result} is therefore a direct NEQ witness. Indeed even at small waiting times we find
	\begin{equation}
		\epsilon_2(\tau_2)= \epsilon_0 +\mathcal J\,\mathcal O\!\left[(\mathcal J\delta_t)^2\right]\label{epsilon_quadratic}
	\end{equation}
	and thus no instantaneous thermalization.
	
	From \eqref{epsilon_quadratic} we notice the evolution is non-Markovian in that it retains memory of $\theta_0$. In the long waiting time limit this memory decays exponentially
	\begin{equation}
		\left|e^{g_{11}(\tau_2,\tau_1)}\right|
		\sim
		\left[\frac{4T_1\theta_1}{\mathcal J}\right]^2
		e^{-\lambda_{L,1}\delta_t},
		\qquad
		\lambda_{L,1}=4T_1\theta_1,
		\label{exp_decay}
	\end{equation}
	with the rate corresponding to the Lyapunov exponent of the thermal reference state given by $\lambda_{L,1}=4T_1\theta_1$ \cite{Maldacena2016Nov,vanmanen}. And so the update rule \eqref{eq:exact_return_result} reduces to the single-quench (from an initial thermal state) rule \eqref{single_quench_result} $\epsilon_2(\tau_2) \xrightarrow[]{\delta_t\to \infty} c_{12}\,\epsilon_1$. With this, subsequent quenches become effectively Markovian, each seeing only the state left by its predecessor.  
	
	Note this Lyapunov-controlled forgetting rate \eqref{exp_decay} is also the \emph{thermalization rate}. Such rates are ordinarily related to scrambling and at low temperatures $\lambda_{L,1}=4T_1\theta_1$, they approach the chaos bound \cite{Maldacena2016Aug} 
	$\lambda_{L,1} \to 2\pi T_1$. This so-called Planckian rate is seen in the thermalization of black holes \cite{Lei2022Apr,Vishveshwara1970Aug,Carullo2021Apr,Louw2023Oct}, numerical SYK studies \cite{Eberlein:2017jb,Osterkorn2026Sep} and via typicality arguments \cite{Goldstein_2015}. Another example is given in this letter. It is analytic and extends beyond the low-temperature regime.

	The expectation value $\Tr[H_0 \rho(\tau_2)]$ is sensitive to the mixed-time solution $e^{g_{01}(t_1,\tau_2)}$, whose magnitude is controlled by $ e^{g_{11}(\tau_1,\tau_2)}=e^{g_1(-\delta_t)}$, where $t_1\in I_0$.
	This resolves the apparent tension between thermal correlation functions and a present NEQ witness. The system behaves thermally at time $\tau_2$ \emph{only} if all correlation functions connecting to $\tau_2$ satisfy the KMS relation. While this is the case for the diagonal blocks, the off-diagonal blocks remain, except for a trivial quench, nonthermal. The decay of the correlator $e^{g_1(-\delta_t)}$ at large $\delta_t$ then removes the non-thermal contribution to the NEQ-witness, and the system becomes effectively indistinguishable from its thermal reference. This makes $\lambda_{L,1}$ the thermalization rate.
	Other equilibrium criteria are studied numerically in \cite{Osterkorn2026Sep} for the same model after a single quench.
	
\begin{figure}[ht]
		\includegraphics[width=\columnwidth]{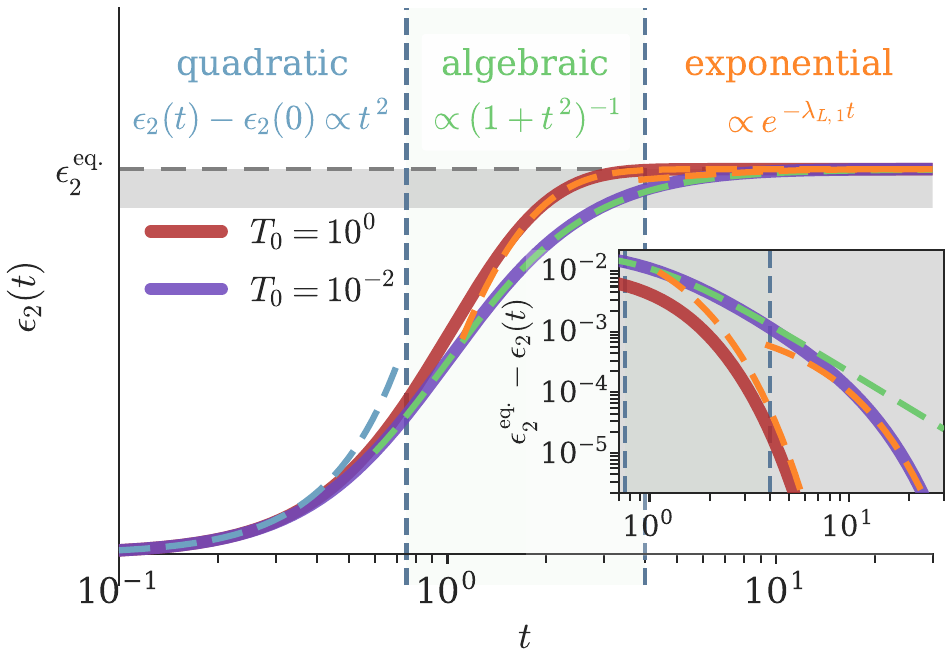}
		\caption{
			Relaxation of the energy density $\epsilon_2(t) \propto \expval{H_0(t)}q^2/N$ towards its post-quench equilibrium value $\epsilon_2^{\rm eq.}$ as the waiting time $t=\delta_t$ before the return quench increases ($\Jj=1$, $c_{01}=c_{12}=0.99$, $\tau_1=0$). Dashed guides mark the three regimes: quadratic \eqref{epsilon_quadratic}
			at early times, algebraic \eqref{eq:lowT_main} at intermediate
			times, and exponential \eqref{exp_decay} at late times, all corresponding to the low-temperature result. The higher temperature result in red has been provided to show fast thermalization relative to the $T_0 = 10^{-2}$ case. 
			The $\epsilon_2$-axis is individually rescaled for comparison. Inset: distance to equilibrium on logarithmic axes, resolving the algebraic-to-exponential crossover. 
		}
		\label{fig:return-quench-probe}
	\end{figure}

	Away from saturation, the thermalization time $\lambda_{L,1}^{-1}$ grows without bound as $T_1\to0$. In this limit an intermediate window opens up, in which the decay is instead controlled by the coupling strength $\mathcal{J}$
	\begin{equation}
		\label{eq:lowT_main}
		\vert e^{g_{11}(\tau_2,\tau_1)}\vert \sim(1+\mathcal J^2\delta_t^2)^{-1},
	\end{equation}
	whose algebraic tail reflects the SYK conformal (IR) regime.
	Thus in the IR, thermalization is initially controlled by $\mathcal{J}$ until it crosses over into the exponential decay around $\delta_t\sim \lambda_{L,1}^{-1}$.
	The full interpolation between the quadratic onset \eqref{epsilon_quadratic} and the algebraic-to-exponential decay \eqref{exp_decay} crossover is shown in Fig.~\ref{fig:return-quench-probe}.
	
	\emph{Entropy witness---.} Lastly, we can associate $\theta_a$ with the coarse-grained entropy, namely the entropy of the corresponding reference thermal state of $ H_a$ whose post-quench correlators coincide, at the order considered. 
	Once the mixed-time memory has decayed, the return protocol cannot distinguish the two states either, giving this association its thermodynamic meaning. 
	
	For a single quench from a thermal state $\Delta S\propto-\Delta(\theta^2)$ \cite{Maldacena2016Nov,Louw2023Oct}. 
	The inequality $\sin\theta_1\leq\sin\theta_0$ obtained from \eqref{single_quench_result} is precisely the ordering expected from the second law for this asymptotic thermodynamic entropy. 
	
	The analogous inequality no longer holds for the second quench. Indeed at finite $\delta_t$ it may increase $\sin\theta_2>\sin\theta_1$ without contradiction, precisely because the intermediate state has not yet thermalized. Subsequent quenches thereby expose the otherwise hidden memory as an operational nonequilibrium witness.
	
	\emph{Conclusion---.} Taken together, \eqref{eq:exact_return_result}, \eqref{epsilon_quadratic} and \eqref{GMrel} show that the probe \eqref{eq:probe} interpolates in real time between the response of the initial
	state and that of the thermal reference state. 
	Two distinct witnesses therefore rule out instantaneous thermalization: the waiting-time dependence of \eqref{eq:probe}, and the entropy decrease that would be inferred if the intermediate state were treated as thermal.
	
	The deviation of \eqref{eq:probe} from the thermal-reference response decays at the Lyapunov rate $\lambda_{L,1}$, giving $\lambda_{L,1}$ an operational meaning as the thermalization rate and rendering subsequent quenches effectively Markovian at late waiting times. As such, the thermality of the post-quench correlator sector does not by itself characterize the state, even though at $N\to\infty$, $q\to\infty$ the fermionic $n$-point functions satisfy KMS exactly on the domain where all time arguments lie after the quench. \\
	
	The return protocol exposes information that is invisible to passive probes of the final-Hamiltonian dynamics but becomes measurable under the subsequent quench. The entropy argument shows that accounting for this information is necessary for a consistent thermodynamic interpretation. 
	
    The IR SYK dynamics admits a holographic interpretation. Each interval of constant Hamiltonian corresponds to a stationary nearly-AdS$_2$ black-hole solution \cite{maldacenastanford2016conformal,Dhar:2018pii, Maldacena2016Nov}, with the ADM mass and Euclidean period encoded in $\theta$. 
    Within each interval the metric and boundary coincide with the
	thermal solution, yet the boundary state can retain nonequilibrium memory. This
	memory is hair-like in the operational sense: invisible to the usual passive
	probes, but sufficient to distinguish the state under subsequent perturbations.
    
    This hidden-memory structure has a close analogue in Kourkoulou--Maldacena states \cite{Kourkoulou:2017zaj,Nosaka:2019tcx}, where a thermal flavor-diagonal correlator sector coexists with state-dependent flavor-off-diagonal correlations.
	In KM states, order $N$ preparation-dependent pair correlations are individually of order one \cite{Kourkoulou:2017zaj}. 
	Here, deviations from the thermal reference are distributed over the $\binom{N}{q}$ $q$-body correlations entering $H_0$:
	they vanish individually at large-$N$, while their coupling-weighted sum remains extensive (see Secs.~\ref{sec:qbody-KM} 
    and~\ref{sec:SM-nonthermal-correlations} of the SM).
    This collective mechanism may also be relevant to
    corpuscular descriptions of black holes%
    ~\cite{Dvali:2012en,Cadoni:2021qfn}.
    
    The instantaneous thermality of the post-quench correlators allows us to isolate the decay of the memory revealed by the return protocol.
    Comparison with finite-$q$ quenches, where these correlators themselves relax toward their thermal form \cite{Bhattacharya2019Jul}, could clarify the interplay between correlator equilibration and memory loss.
    The finite-$q$ return response for weak quenches,
    Eq.~\eqref{eq:SM-weak-return-epsilon} of the SM,
    provides a starting point for this comparison.

    \emph{Outlook---.} The behavior of the nonequilibrium thermodynamic entropy during the waiting time itself remains an open question, since $\theta$ only tracks the thermodynamic entropy asymptotically. When viewed from a subsystem perspective $\theta$ can be related to the entanglement entropy \cite{Zhang2022Aug}. As such, it need not approach its asymptotic value monotonically, and nothing forbids an overshoot followed by relaxation back down. This question could be explored through studying the subsystem entropy either analytically or numerically \cite{Zhang:2020iep}.
	
	\begin{acknowledgments}
		\emph{Acknowledgments---.} This work was done as part of the MSc of J.L. at the Faculty of Physics of Ludwig Maximilian University of Munich. J.L. thanks Niklas Ziereis for useful discussions.
	\end{acknowledgments}
	
	\bibliography{references}
	\appendix
    \setcounter{secnumdepth}{2}
	\title{Supplementary material of "Lyapunov-controlled thermalization: an exact real-time example"}

\section{Conventions and Kadanoff-Baym equations}
\label{sec:SM-conventions}

Throughout the paper, we follow the large-$q$ SYK conventions of Refs.~\cite{Maldacena2016Nov,Eberlein:2017jb}. 
The Hamiltonian \eqref{the hamiltonian} is a sum
of SYK Hamiltonians with mutually uncorrelated
disorder realizations,
\begin{align}
    H(t)
    &=\sum_{j=1}^{L}
    H^{(j)}[\mathcal J^{(j)}(t)]
    \label{eq:SM-Hamiltonian}\\
    &=
    \imath^{q/2}\sum_I j_I(t)\psi_I,
    \qquad
    \psi_I\equiv\psi_{i_1}\cdots\psi_{i_q},
    \notag
\end{align}
with multiindices $I=(i_1<\cdots<i_q)$.
The Majorana fermions obey
\(\{\psi_\mu,\psi_\nu\}=\delta_{\mu\nu}\).
The total random couplings $j_I(t)$ are
Gaussian with zero mean; denoting the disorder
average by an overbar, their covariance is
\begin{align}
    \overline{j_I(t_1)j_J(t_2)}
    &=
    \delta_{IJ}
    \frac{(q-1)!}{N^{q-1}}
    \frac{2^{q-1}}{q}\,
    \vec{\mathcal J}(t_1)\cdot
    \vec{\mathcal J}(t_2),
    \label{eq:SM-covariance}
\end{align}
where the scalar product denotes the sum
$\sum_{j=1}^{L}
\mathcal J^{(j)}(t_1)\mathcal J^{(j)}(t_2)$
over the SYK components.
For each component $H^{(j)}$, the disorder
realization is fixed throughout the protocol;
only its effective coupling strength
$\mathcal J^{(j)}(t)$ changes.
Thus, the protocol enters the disorder-averaged
theory only through the scalar product
$\vec{\mathcal J}(t_1)\cdot\vec{\mathcal J}(t_2)$.

At the large-$N$ saddle, we define
\begin{align}
    G^{>}(t_1,t_2)
    &\equiv
    -\frac{\imath}{N}
    \sum_{\mu=1}^{N}
    \overline{
    \Tr\!\left[
        \rho_0\,
        \psi_\mu(t_1)\psi_\mu(t_2)
    \right]},
    \label{eq:SM-Ggreater}
\end{align}
where \(\rho_0=e^{-\beta_0H_0}/\Tr e^{-\beta_0H_0}\) is the initial thermal density matrix.
The greater and lesser Green functions are related by
\begin{align}
    G^{<}(t_1,t_2)
    &=
    \bigl[G^{>}(t_1,t_2)\bigr]^*
    =
    -G^{>}(t_2,t_1).
    \label{eq:SM-Glesser}
\end{align}
For the covariance \eqref{eq:SM-covariance}, the greater and lesser self-energies are
\begin{align}
    \Sigma^{\gtrless}(t_1,t_2)
    &=
    -\imath^q
    \frac{2^{q-1}}{q}\,
    \vec{\mathcal J}(t_1)\cdot
    \vec{\mathcal J}(t_2)\,
    \bigl[G^{\gtrless}(t_1,t_2)\bigr]^{q-1}.
    \label{eq:SM-self-energy-finite-q}
\end{align}
For \(X\in \{G,\Sigma\}\), we define 
\begin{align}
    X^{R}(t_1,t_2)
    &=
    \Theta(t_1-t_2)
    \bigl[
        X^{>}(t_1,t_2)-X^{<}(t_1,t_2)
    \bigr],
    \nonumber\\
    X^{A}(t_1,t_2)
    &=
    -\Theta(t_2-t_1)
    \bigl[
        X^{>}(t_1,t_2)-X^{<}(t_1,t_2)
    \bigr].
    \label{eq:SM-retarded-advanced}
\end{align}
Before taking the large-$q$ limit, the KBE takes the standard form \cite{Eberlein:2017jb}
\begin{align}
    \imath\partial_{t_1}G^{>}(t_1,t_2)
    &=
    \int_{-\infty}^{\infty}\!dt_3\,
    \Sigma^{R}(t_1,t_3)G^{>}(t_3,t_2)
    \nonumber\\
    &\quad+
    \int_{-\infty}^{\infty}\!dt_3\,
    \Sigma^{>}(t_1,t_3)G^{A}(t_3,t_2).
    \label{eq:SM-KBE-finite-q}
\end{align}
The companion Kadanoff-Baym equation for the second time argument follows by complex conjugation and \(t_1\leftrightarrow t_2\), and is therefore not independent.
Thus, the full time dependence of the protocol enters the Kadanoff-Baym equation through the self-energy \eqref{eq:SM-self-energy-finite-q}, and hence through \(\vec{\mathcal J}(t_1)\cdot\vec{\mathcal J}(t_2)\).
We denote the evolving density matrix by $\rho(t)$.
For arbitrary even $q$, the normalized energy
$\epsilon(t)\equiv 2q^2\Tr[H(t)\rho(t)]/N$
fixes the equal-time boundary condition of
Eq.~\eqref{eq:SM-KBE-finite-q}
through the Galitskii--Migdal relation,
\begin{align}
 q
 \partial_{t}\ln\!\bigl[2\imath G^{>}(t,t_2)\bigr]
 \Bigr|_{t_2=t}
 &=
 \imath\epsilon(t)
 =
 2\imath\frac{ q^2}{N}\Tr[H(t)\rho(t)].
 \label{eq:SM-energy-finite-q}
\end{align}
The standard large-$q$ parametrization is
\begin{align}
    G^{>}(t_1,t_2)
    &=
    -\frac{\imath}{2}
    \exp\!\left[
        \frac{1}{q}g(t_1,t_2)
        +\mathcal O(q^{-2})
    \right].
    \label{eq:SM-large-q-parametrization}
\end{align}
Thus, from \(G^>(t,t)=-\imath/2\) and \eqref{eq:SM-Ggreater},
\begin{align}
    g(t,t)&=0,
    &
    g(t_2,t_1)&=g(t_1,t_2)^*.
    \label{eq:SM-g-properties}
\end{align}
The self-energy \eqref{eq:SM-self-energy-finite-q} reduces to
\begin{align}
    \label{eq:SM-self-energy-large-q}
    \Sigma^{>}(t_1,t_2)
    &=
    -\frac{\imath}{q}\,
    \vec{\mathcal J}(t_1)\cdot
    \vec{\mathcal J}(t_2)\,
    e^{g(t_1,t_2)}
    +\mathcal O(q^{-2}),\\
    \Sigma^{<}(t_1,t_2)
    &=
    -\Sigma^{>}(t_2,t_1).\nonumber
\end{align}
Substituting \eqref{eq:SM-large-q-parametrization} and \eqref{eq:SM-self-energy-large-q} into the Kadanoff-Baym equations gives, for \(t_2\geq t_1\),
\begin{align}
    \partial_{t_1}g(t_1,t_2)
    &=
    \imath\epsilon(t_1)
    +
    2\int_{t_1}^{t_2}\!dt_3\,
    \vec{\mathcal J}(t_1)\cdot
    \vec{\mathcal J}(t_3)\,
    e^{g(t_1,t_3)},
    \label{eq:SM-large-q-KBE}\\
    \epsilon(t)
    &=
    \Im\!\left[
        \int_{-\infty}^{t}\!dt_3\,
        2\vec{\mathcal J}(t)\cdot
        \vec{\mathcal J}(t_3)\,
        e^{g(t,t_3)}
    \right].
\label{eq:SM-epsilon}
\end{align}
At equal times,
\begin{align}
    \left.
    \partial_{t_1}g(t_1,t_2)
    \right|_{t_2=t_1^+}
    &=
    \imath\,\epsilon(t_1)
    =
    \frac{2\imath q^2}{N}
    \Tr[H(t_1)\rho(t_1)],
    \label{eq:SM-equal-time-boundary}
\end{align}
where the second equality follows from the Galitskii-Migdal relation.
Differentiating \eqref{eq:SM-large-q-KBE} with respect to \(t_2\) further gives
\begin{align}
    \partial_{t_1}\partial_{t_2}g(t_1,t_2)
    &=
    2\,
    \vec{\mathcal J}(t_1)\cdot
    \vec{\mathcal J}(t_2)\,
    e^{g(t_1,t_2)},
    \label{eq:SM-Liouville}
\end{align}
which is the Liouville-type equation used in the main text.

\newpage

\section{Construction of the multiquench solution}
\label{app:multiquench-derivation}

In the main text we solved the Kadanoff-Baym equation (KBE) for a single return
quench. Here we construct the solution for an arbitrary sequence of quenches,
and show that it is the unique regular solution of the large-$q$ equations of
motion. The strategy is a block-wise recursion: each block of the two-time
plane is solved in closed form in terms of the two blocks lying closer to the
diagonal, and the diagonal blocks themselves are fixed by a single
self-consistency condition.

\begin{figure}[ht]
\centering
\begin{tikzpicture}[scale=1]
    \def\T{1.5}
    \def\Xmax{4.2}

    % Orange cross highlight around cell (1,1)
    \fill[orange!20] (0*\T,1*\T) rectangle (1*\T,2*\T);   % (1,2)
    \fill[orange!20] (0*\T,0*\T) rectangle (1*\T,1*\T);   % (1,1)
    \fill[orange!20] (1*\T,0*\T) rectangle (2*\T,1*\T);   % (2,1)
    \fill[orange!20] (-1*\T,0*\T) rectangle (0*\T,1*\T);  % (0,1)
    \fill[orange!20] (0*\T,-1*\T) rectangle (1*\T,0*\T);  % (1,0)

    % Diagonal line
    \draw[black!30, thick, dashed] (-1*\T,-1*\T) -- ({\Xmax*\T},{\Xmax*\T});

    % Grid boundaries
    \foreach \x in {0, 1, 2, 3} {
        \draw[RoyalBlue!90, line width=1pt] (\x*\T,-1*\T) -- (\x*\T,{\Xmax*\T});
        \draw[RoyalBlue!90, line width=1pt] (-1*\T,\x*\T) -- ({\Xmax*\T},\x*\T);
    }

    % Concrete cell labels
    \foreach \a/\x in {0/-0.5, 1/0.5, 2/1.5, N/3.6} {
        \foreach \b/\y in {0/-0.5, 1/0.5, 2/1.5, N/3.6} {
            \node[fill=white, fill opacity=0.4, text opacity=1, inner sep=2pt] 
                at ({\x*\T},{\y*\T}) {$(\a,\b)$};
        }
    }

    % Ellipses for grid continuation
    \foreach \y in {-0.5, 0.5, 1.5, 3.6} {
        \node at (2.5*\T, {\y*\T}) {$\dots$};
    }
    \foreach \x in {-0.5, 0.5, 1.5, 3.6} {
        \node at ({\x*\T}, 2.5*\T) {$\dots$};
    }
    \node at (2.5*\T, 2.5*\T) {$\ddots$};

    % Arrows into (1,2)
    \draw[line width=3pt, -{Stealth[length=14pt, open, round]} , StrangeGreen!90] (0.5*\T,0.8*\T) -- (0.5*\T,1.25*\T);
    \draw[line width=3pt, -{Stealth[length=14pt, open, round]} , StrangeGreen!90] (1.2*\T,1.5*\T) -- (0.88*\T,1.5*\T);
    %   (0.5*\T,0.8*\T) -- (0.5*\T,1.25*\T);
    %\draw[line width=2pt, -{Stealth[length=14pt, open, round]} , StrangeGreen!90] (1.3*\T,1.5*\T) -- (0.88*\T,1.5*\T);
    \node[fill=white, fill opacity=0.8, text opacity=1, inner sep=2pt, text = StrangeGreen!90] at (0.5*\T, 2.2*\T) {\textbf{recursion}};
    %\node[fill=white, inner sep=3pt, StrangeGreen!90] at (0.6*\T, 2.2*\T) {\textbf{recursion}};

    % Axis tick labels
    \node[below] at (0*\T, -1*\T) {$\tau_1$};
    \node[below] at (1*\T, -1*\T) {$\tau_2$};
    \node[below] at (2*\T, -1*\T) {$\tau_3$};
    \node[below] at (2.5*\T, -1*\T) {$\dots$};
    \node[below] at (3*\T, -1*\T) {$\tau_N$};

    \node[left] at (-1*\T, 0*\T) {$\tau_1$};
    \node[left] at (-1*\T, 1*\T) {$\tau_2$};
    \node[left] at (-1*\T, 2*\T) {$\tau_3$};
    \node[left] at (-1*\T, 2.5*\T) {$\vdots$};
    \node[left] at (-1*\T, 3*\T) {$\tau_N$};

        % Axes
    \draw[->, thick] (-1*\T,-1*\T) -- ({\Xmax*\T},-1*\T) node[right] {$t_1$};
    \draw[->, thick] (-1*\T,-1*\T) -- (-1*\T,{\Xmax*\T}) node[above] {$t_2$};
\end{tikzpicture}
\caption{Partition of the $(t_1,t_2)$-plane into blocks $I_a \times I_b$. The continuity boundary conditions \eqref{eq:multiquench-matching} are imposed on the blue connecting lines. The green arrows illustrate the recursive propagation of boundary data.}
\label{fig:grid2}
\end{figure}
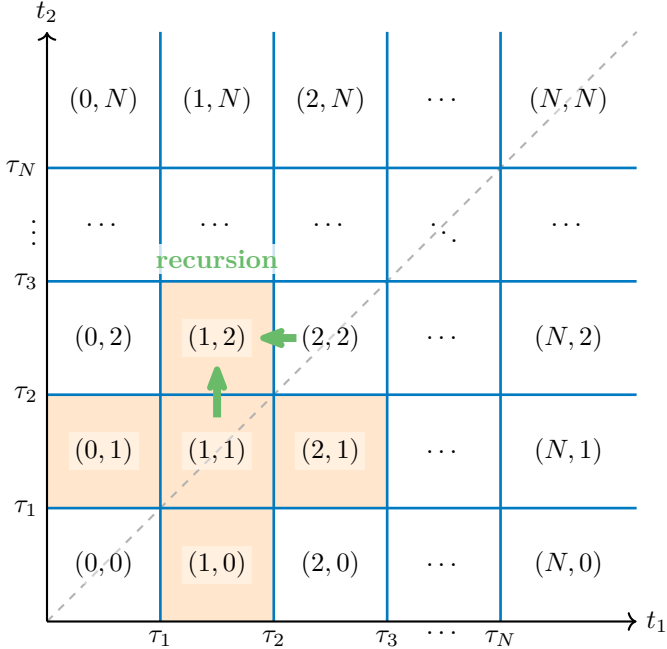

Let the couplings be piecewise constant on the intervals
$I_a=(\tau_a,\tau_{a+1})$, with $\tau_0=-\infty$,
as shown in Fig.~\ref{fig:grid2}.
Continuity of
the two-point function across the quench lines therefore requires the
\emph{matching conditions}
\begin{align}
    g_{a,b}(t_1,\tau_b)
    &= g_{a,b-1}(t_1,\tau_b) \notag\\
    g_{a,b}(\tau_{a+1},t_2)
    &= g_{a+1,b}(\tau_{a+1},t_2).
    \label{eq:multiquench-matching}
\end{align}

For the general protocol, define $\mathcal J_a\equiv|\vec{\mathcal J}_a|$ 
and $c_{ab}\equiv\vec{\mathcal J}_a\cdot\vec{\mathcal J}_b/(\mathcal J_a\mathcal J_b)$.
We collect the quench strengths into
\begin{align}
    K_{ab}^2 \equiv \vec{\mathcal J}_a\cdot\vec{\mathcal J}_b .
\end{align}
On every block $I_a\times I_b$ the covariance $K_{ab}^2$ is constant, hence
the large-$q$ equation of motion reduces to a Liouville equation
\begin{align}
\label{eq:regionwise-liouville}
    \partial_{t_1}\partial_{t_2}g_{ab}(t_1,t_2)
    = 2 \vec{\mathcal J}_a\cdot\vec{\mathcal J}_b\,e^{g_{ab}(t_1,t_2)},\; (t_1,t_2) \in I_a \times I_b
\end{align}
We restrict throughout to $a \leq b$, corresponding to $t_1\leq t_2$; the
remaining half of the plane follows from $g(t_2,t_1)=g(t_1,t_2)^*$. Its two edges closest to the equal-time line are $t_2=\tau_b$ and $t_1=\tau_{a+1}$.
Integrating \eqref{eq:regionwise-liouville} over the rectangle with corners $(\tau_{a+1},\tau_b)$ and $(t_1,t_2)$ we are left with the integral equation
\begin{align}
    g_{ab}(t_1,t_2)
    =&\, g_{a,b-1}(t_1,\tau_b) + g_{a+1,b}(\tau_{a+1},t_2) \notag\\
    &- \ln\Lambda_{ab}- 2\ln D_{ab}(t_1,t_2).\label{eq:multiquench-region-solution}
\end{align}
where we have defined
\begin{equation}
    \ln D_{ab}(t_1,t_2) =
    \vec{\mathcal J}_a\cdot\vec{\mathcal J}_b
    \int_{t_1}^{\tau_{a+1}}\!\!ds_1
    \int_{\tau_b}^{t_2}\!\!ds_2\,
    e^{g_{ab}(s_1,s_2)} \label{eq:multiquench-integral-equation}
\end{equation}
as well as the corner term
\begin{align}
\label{eq:corner-abbrev}
    \Lambda_{ab}
    \equiv 
    e^{g_{a+1,b}(\tau_{a+1},\tau_b)}.
\end{align}

\subsection{The block solution}

\begin{proposition}
\label{prop:block}
Let $a<b$ and suppose the blocks $(a,b-1)$ and $(a+1,b)$ are known and satisfy
\eqref{eq:multiquench-matching}. Then we claim that the unique solution to \eqref{eq:regionwise-liouville}, with matching conditions \eqref{eq:multiquench-matching}, is \eqref{eq:multiquench-region-solution} with the function
\begin{equation}
    D_{ab}(t_1,t_2) = 1-\frac{ c_{ab}\, V_{a,b-1}(t_1)\, U_{a+1,b}(t_2)}{\Lambda_{ab}}, \label{Dsol}
\end{equation}
where we have defined the boundary integrals: for $t\in I_b$ and
$t\in I_a$ respectively
\begin{align}
\label{eq:multiquench-UV-definition}
    U_{a,b}(t) &\equiv \Jj_{b}
    \int_{\tau_b}^{t}\!dt_3\, e^{g_{a,b}(\tau_a,t_3)}\\
    V_{a,b}(t) &\equiv \Jj_{a}
    \int_{\tau_{a+1}}^{t}\!dt_3\, e^{g_{a,b}(t_3,\tau_{b+1})}.\notag
\end{align}
Note that $U_{a,b}$ propagates boundary data along the left edge of the block and $V_{a,b}$ along its upper edge.

\end{proposition}

\paragraph*{The equation of motion.}

The mixed derivative of \eqref{eq:multiquench-region-solution} comes entirely from the last term $-2\ln D_{ab}$. Using
\begin{align}
    \partial_{t_1}\partial_{t_2}\ln D_{ab}
    =
    \frac{
        D_{ab}\,D_{ab}^{(1,1)}
        -
        D_{ab}^{(1,0)}\,D_{ab}^{(0,1)}
    }{D_{ab}^2},
    \label{ddD}
\end{align}
where we denote by the superscript $D^{(i,j)}$ the $i$'th derivative of the first argument and the $j$'th derivative of the second.
From the definitions \eqref{eq:multiquench-UV-definition} we have $U_{a,b}'(t) = \Jj_{b} e^{g_{a,b}(\tau_a,t)},$
$V_{a,b}'(t)= \Jj_{a} e^{g_{a,b}(t,\tau_{b+1})},$ which implies
\begin{align}
    D_{ab}^{(1,1)}(t_1,t_2)
    &=
    -\frac{c_{ab}\Jj_{a}\Jj_{b}}{\Lambda_{ab}}
    e^{g_{a,b-1}(t_1,\tau_b)+g_{a+1,b}(\tau_{a+1},t_2)} \notag\\
    D_{ab}^{(1,0)}(t_1,t_2)
    &=
    -\frac{c_{ab}\Jj_{a}}{\Lambda_{ab}}
    e^{g_{a,b-1}(t_1,\tau_b)}U_{a+1,b}(t_2) \notag\\
    D_{ab}^{(0,1)}(t_1,t_2)
    &=
    -\frac{c_{ab}\Jj_{b}}{\Lambda_{ab}}
    V_{a,b-1}(t_1)e^{g_{a+1,b}(\tau_{a+1},t_2)}.
\end{align}
The two contributions in the numerator of \eqref{ddD} combine, the second cancelling the term
proportional to $(1-D_{ab})$ in the first, leaving only
\begin{align}
    \partial_{t_1}\partial_{t_2}g_{ab}
    =&
    2K_{ab}^2\,
    \frac{
        e^{
            g_{a,b-1}(t_1,\tau_b)
            +
            g_{a+1,b}(\tau_{a+1},t_2)
        }
    }{
        \Lambda_{ab}D_{ab}^2
    }
\end{align}
which reduces to $2K_{ab}^2e^{g_{ab}(t_1,t_2)}$; hence \eqref{eq:multiquench-region-solution} solves the Liouville equation
inside $I_a\times I_b$. What remains to be shown is that this solution is unique and that the boundary conditions are satisfied.

\paragraph*{The lower edge.}
At $t_2=\tau_b$ we have $U_{a+1,b}(\tau_b)=0$ read off from \eqref{eq:multiquench-UV-definition}, hence $D_{ab}=1$. The remaining
$t_2$-dependent term in \eqref{eq:multiquench-region-solution} becomes
$g_{a+1,b}(\tau_{a+1},\tau_b)=\ln\Lambda_{ab}$, which cancels the third term
identically. We are therefore left with the first matching condition \eqref{eq:multiquench-matching} $g_{ab}= g_{a,b-1}$ at $t_2 = \tau_b$.

\paragraph*{The right edge.}
At $t_1=\tau_{a+1}$ we have $V_{a,b-1}(\tau_{a+1})=0$, hence again $D_{ab}=1$, and \eqref{eq:multiquench-region-solution} reduces to 
$g_{ab}(\tau_{a+1},t_2)= g_{a+1,b}(\tau_{a+1},t_2),$ since the final terms 
$g_{a,b-1}(\tau_{a+1},\tau_b)- g_{a+1,b}(\tau_{a+1},\tau_b)$ cancel due to the premise of the proposition, namely that blocks $(a,b-1)$ and $(a+1,b)$ satisfy \eqref{eq:multiquench-matching}. Together we have thus proven that \eqref{Dsol} is a solution but not yet that it is unique.

\emph{Uniqueness.} Since the nonlinearity $e^{g}$ is locally Lipschitz in $g$, Picard iteration of
\eqref{eq:multiquench-integral-equation} converges on any subrectangle on
which the iterates stay bounded, and the fixed point is unique. Here we call  a 
solution \emph{regular} when it is continuous and has no singularities at finite times, which for \eqref{eq:multiquench-region-solution} amounts to $D_{ab}\neq0$; the zeros of $D_{ab}$ are the usual movable singularities of
the Liouville equation. As such \eqref{eq:multiquench-region-solution} is the complete regular solution on
$I_a\times I_b$.

\subsection{Recursive solution}
\label{SM-recursiveresolution}

The recursion of Proposition~\ref{prop:block} is anchored on the diagonal,
where the solution retains the single-quench form
\begin{align}
\label{eq:multiquench-diagonal-solution}
    e^{g_{aa}(t_1,t_2)}
    =
    \frac{
        \cos^2\theta_a
    }{
        \cosh^2\!\left[
            \tfrac{\lambda_{L, a }}{2}(t_2-t_1)-\imath\theta_a
        \right]
    }
\end{align}
\begin{equation}
    \lambda_{L, a }=2\mathcal J_a\cos\theta_a,\quad \epsilon_a=-2\mathcal J_a\sin\theta_a .
\end{equation}

Inserting this into \eqref{eq:multiquench-UV-definition} and using
$\int dx\,\mathrm{sech}^2(x)=\tanh(x)$, one may check that
\begin{align}
\label{eq:multiquench-diagonal-UV}
    U_{a, a }(t) &=\frac{2 \Jj_{a}}{\lambda_{L, a }\coth\!\left[\lambda_{L, a }(t-\tau_{a})/2\right]+\imath \epsilon_a}\\
    V_{a, a }(t) &= \frac{2\Jj_{a}}{\lambda_{L, a }\coth\!\left[\lambda_{L, a }(t-\tau_{a+1})/2\right]-\imath \epsilon_a},\notag
\end{align}
where the addition formula $\tanh(x-\imath\theta_a)$, together with
$\tanh(\imath\theta_a)=\imath\tan\theta_a$ and
$\imath \epsilon_a=-\imath\lambda_{L, a }\tan\theta_a$, has been used to bring the result
into this compact form.

The functions $U$ and $V$ entering \eqref{eq:multiquench-region-solution} need not be recomputed by quadrature at
each step; they satisfy a closed recursion of their own. Setting
$t_1=\tau_{a}$ in \eqref{eq:multiquench-region-solution}, the integrand of
$U_{a,b}$ becomes a perfect differential in the variable $U_{a+1,b}$, since
$U^{\prime}_{a+1,b}(t)=\Jj_{b}e^{g_{a+1,b}(\tau_{a+1},t)}$. The elementary integral
$\int_0^{u}du'(1-\alpha u')^{-2}=u/(1-\alpha u)$ then yields, and the mirror
computation at $t_2=\tau_{b+1}$ likewise yields,
\begin{align}
\label{eq:multiquench-UV-recursion}
    U_{a,b}(t)&=\frac{U_{a+1,b}(t)\,V^{\prime}_{a,b-1}(\tau_a)}{ V^{\prime}_{a,b-1}(\tau_{a+1})  -c_{ab}\Jj_a\,U_{a+1,b}(t)\,V_{a,b-1}(\tau_a)},\\
    V_{a,b}(t)&=\frac{ V_{a,b-1}(t)\,U^{\prime}_{a+1,b}(\tau_{b+1})}{U^{\prime}_{a+1,b}(\tau_b) -c_{ab}\Jj_b\,V_{a,b-1}(t)\,U_{a+1,b}(\tau_{b+1})}.\notag
\end{align}
Note that $V^{\prime}_{a,b-1}(\tau_{a+1})/\Jj_a
=e^{g_{a,b-1}(\tau_{a+1},\tau_b)}$, which equals
$\Lambda_{ab}$. As such the corner
identity is what allows the two recursions to be written in the symmetric form
above.
\subsection{Self-consistency and the algorithm}
\label{SMss:self-con}

Suppose the solution has been constructed up to the $n$th interval. The
equal-time boundary condition fixes the new diagonal block through
\begin{align}
\label{eq:multiquench-self-consistency}
    \epsilon_{n+1}
    = 2\Jj_{n+1}\operatorname{Im}\left[\sum_{j=0}^{n} c_{j,n+1}\, V_{j,n}(\tau_{j}) \right],
\end{align}
which follows from evaluating the memory integral of the main text at
$t=\tau_{n+1}^+$ and using $g(t,t_3)=g(t_3,t)^*$ to orient every contribution along
the $V$ direction. Every quantity on the right-hand side is already known,
hence \eqref{eq:multiquench-self-consistency} determines $\epsilon_{n+1}$, and
with it $\theta_{n+1}$ and the diagonal block $g_{n+1,n+1}$.
Proposition~\ref{prop:block} then determines
\begin{align}
    g_{n,n+1},\quad
    g_{n-1,n+1},\quad
    \ldots,\quad
    g_{0,n+1}
\end{align}
in turn, each from the two blocks lying one step closer to the diagonal, while
\eqref{eq:multiquench-UV-recursion} updates the boundary integrals alongside.
Thus the diagonal self-consistency condition together with the block-wise
recursion determines the complete upper half of the two-time plane.

Lastly, we record why the second-order equation
\eqref{eq:regionwise-liouville} carries no less information than 
the companion first-order KBE for the second time argument. Differentiating this equation with respect to $t_1$ discards a function of $t_2$, which is restored by the one-sided
diagonal condition
\begin{align}
\label{eq:multiquench-diagonal-boundary}
    \lim_{t_1\to t_2}
    \partial_{t_2}g(t_1,t_2)
    =
    -\imath\epsilon(t_2).
\end{align}
This follows from evaluating the KBE at equal times, where the interval $[t_1,t_2]$ collapses as $t_1\to t_2^-$.
As such \eqref{eq:regionwise-liouville} together with \eqref{eq:multiquench-diagonal-boundary} is equivalent to the original first-order equation.

\subsection{Worked example: Calculating the off-diagonal}
\label{sec:example}

The recursion is easiest to follow on a concrete case. Suppose the first two
intervals have been solved, so that we know the three blocks
\begin{align}
    g_{00},\quad g_{01},\quad g_{11},
\end{align}
together with their boundary integrals $U_{a,b}$ and $V_{a,b}$ for the same 
three index pairs. We now add the interval $I_2$, which means filling the
column $b=2$, namely the blocks $g_{22}$, $g_{12}$ and $g_{02}$.
Here, rows, columns, and parent directions refer to the matrix
$(g_{ab})$, with $a$ the row index and $b$ the column index.

Before starting it is worth stating the bookkeeping rule that governs every
step. Reading off \eqref{eq:multiquench-region-solution} and \eqref{Dsol},
the block $(a,b)$ consumes exactly four objects:
\begin{align}
\label{eq:ingredient-rule}
    \underbrace{g_{a,b-1},\ V_{a,b-1}}_{\text{left parent}},
    \qquad
    \underbrace{g_{a+1,b},\ U_{a+1,b}}_{\text{lower parent}} .
\end{align}
As such it always draws its $V$ from the block one step to the left and its
$U$ from the block one step below, as indicated by the green arrows in
Fig.~\ref{fig:grid2}. 

The corner term $\Lambda_{ab}$ is likewise supplied by
the lower parent alone. Starting from the diagonal and moving up the column at
fixed $b$, both parents are therefore always already in hand.

\paragraph*{Step $0$: seeding the diagonal.}
The new diagonal block is not obtained from \eqref{eq:multiquench-region-solution}
but from the self-consistency condition
\eqref{eq:multiquench-self-consistency}. At $n=1$ it reads
\begin{align}
    \epsilon_2
    = 2\Jj_{2} \operatorname{Im}\left[
    c_{02} V_{0,1}(\tau_0)+ c_{12} V_{1,1}(\tau_1)\right],
\end{align}
whose right-hand side involves only the two known blocks $(0,1)$ and $(1,1)$.
This fixes $\epsilon_2$, hence $\theta_2$, hence $g_{22}$ through
\eqref{eq:multiquench-diagonal-solution}. Its boundary integrals $U_{2,2}$ and
$V_{2,2}$ then follow in closed form from
\eqref{eq:multiquench-diagonal-UV}, with no quadrature required. 

\paragraph*{Step $1$: the block $(1,2)$.}
Here the left parent is $(1,1)$ and the lower parent is $(2,2)$, so both are
diagonal. The corner term is the value of the lower parent at equal times,
\begin{align}
    \Lambda_{12}
    =
    e^{g_{22}(\tau_2,\tau_2)}
    =
    1,
\end{align}
by the boundary condition $g(t,t)=0$. Substituting the rule
\eqref{eq:ingredient-rule} into \eqref{eq:multiquench-region-solution} and
\eqref{Dsol} we are left with
\begin{align}
\label{eq:example-g12}
    g_{12}(t_1,t_2)
    &=g_{11}(t_1,\tau_2)+g_{22}(\tau_2,t_2)- 2\ln D_{12}(t_1,t_2),
    \\
    D_{12}(t_1,t_2)
    &= 1-c_{12}\,V_{1,1}(t_1)\, U_{2,2}(t_2),
\end{align}
for $(t_1,t_2)\in I_1\times I_2$. Every object on the right-hand side was
produced in Step~$0$ or earlier. 
One may check the two matching conditions directly:
at $t_2=\tau_2$ we have $U_{2,2}(\tau_2)=0$ and
$g_{22}(\tau_2,\tau_2)=0$, leaving $g_{12}=g_{11}$;
at $t_1=\tau_2$ we have $V_{1,1}(\tau_2)=0$,
leaving $g_{12}=g_{22}$.

To continue up the column we need the boundary integrals of the block just
constructed. Rather than integrating \eqref{eq:example-g12}, we use the
recursion \eqref{eq:multiquench-UV-recursion} at $(a,b)=(1,2)$,
\begin{align}
\label{eq:example-UV12}
    U_{1,2}(t)&=\frac{U_{2,2}(t)\,V^{\prime}_{1,1}(\tau_1)}{V^{\prime}_{1,1}(\tau_2) -c_{12}\Jj_1 U_{2,2}(t)\,V_{1,1}(\tau_1)},\\
    V_{1,2}(t)&=\frac{V_{1,1}(t)\,U^{\prime}_{2,2}(\tau_3)}{U^{\prime}_{2,2}(\tau_2) -c_{12}\Jj_2\,V_{1,1}(t)\,U_{2,2}(\tau_3)}.
\end{align}
Note that $U_{1,2}$ is what Step~$2$ will consume, whereas $V_{1,2}$ is not
needed until the column $b=3$. Both are rational functions of quantities
already available, hence the update costs nothing beyond an evaluation.

\paragraph*{Step $2$: the block $(0,2)$.}
Neither parent is diagonal now. The left parent is $(0,1)$, known from the
previous column, and the lower parent is $(1,2)$, constructed in Step~$1$. The
corner term is
\begin{align}
    \Lambda_{02} = e^{g_{12}(\tau_1,\tau_2)},
\end{align}
which is \eqref{eq:example-g12} evaluated at the lower left corner of the
block $(1,2)$ and is therefore no longer unity. The same substitution as
before yields
\begin{align}
    g_{02}(t_1,t_2)
    ={}& g_{01}(t_1,\tau_2)+ g_{12}(\tau_1,t_2)\notag\\
    &-\ln\Lambda_{02}- 2\ln D_{02}(t_1,t_2),
    \\
    D_{02}(t_1,t_2)={}&
    1-\frac{c_{02}\,V_{0,1}(t_1)\,U_{1,2}(t_2)}{\Lambda_{02}},
\end{align}
for $(t_1,t_2)\in I_0\times I_2$. This completes the column,
\begin{align}
    (2,2)\;\longrightarrow\;(1,2)\;\longrightarrow\;(0,2),
\end{align}
and applying \eqref{eq:multiquench-UV-recursion} once more at
$(a,b)=(0,2)$ supplies $U_{0,2}$ and $V_{0,2}$ for the next quench.

In Step~$2$, both parents are off-diagonal, whereas in Step~$1$ both
are diagonal. Every subsequent
step in a longer column is of the Step~$2$ type, hence the two steps above
already exhibit the general pattern. Further, at no point is a block or a
boundary integral required before it has been constructed, so the procedure
closes.
\newpage

\section{Relation to Kourkoulou--Maldacena states}
\label{sec:KM-qbody-relation}
Ref.~\cite{Kourkoulou:2017zaj} introduced the
Kourkoulou--Maldacena (KM) states, which are pure states
in the SYK model.
After preparation in a KM state, the system evolves
under the SYK Hamiltonian $H_{\mathrm{SYK}}$.
At leading large $N$, the system appears thermal
in terms of a large class of correlation functions,
yet retains memory of the initial preparation.
This memory provides a nonequilibrium witness
reminiscent of Eq.~\eqref{eq:probe} in the main text.
We review these established results in
Sec.~\ref{sec:KM-review}, while in
Sec.~\ref{sec:qbody-KM} we construct a $q$-body analogue
for comparison with the quenched states and their
corresponding nonequilibrium witness,
Eq.~\eqref{eq:exact_return_result}.

\subsection{Review of the KM states}
\label{sec:KM-review}
The KM basis states are defined by 
\begin{align}
    S_k=-2\imath\psi_{2k-1}\psi_{2k},\quad S_k|B_s\rangle=  s_k|B_s\rangle,  \quad  s_k=\pm1,
\end{align}
where each of the \(2^{N/2}\) possible choices of signs
\(s\equiv(s_1,\ldots,s_{N/2})\) corresponds to one
particular KM basis state \(|B_s\rangle\).
Together, these states form a complete orthonormal basis.
Ref.~\cite{Kourkoulou:2017zaj} studies KM states at
inverse temperature \(\beta=1/T\), prepared by
Euclidean evolution with the SYK Hamiltonian
\(H_{\mathrm{SYK}}\),
\(|B_s(\beta)\rangle\propto
e^{-\beta H_{\mathrm{SYK}}/2}|B_s\rangle\).

Here, we focus on $\beta=0$, where the
KM states coincide with the basis states $|B_s\rangle$.
The subsequent real-time evolution is 
generated by \(H_{\mathrm{SYK}}\).
From the completeness of the basis we have
\begin{align}
    2^{-N/2}\sum_s
    \langle B_s|\mathcal O|B_s\rangle
    &=2^{-N/2}\Tr\mathcal O
    =\langle\mathcal O\rangle_{\beta=0},
\label{eq:KM-completeness}
\end{align}
for any operator \(\mathcal O\). 

The disorder-averaged SYK model has an \(O(N)\) symmetry
in the large-\(N\) limit. 
Under simultaneous \(O(N)\) transformations of the fermions
\(\psi_i\) and the coupling tensor \(j_I\), the Hamiltonian
remains invariant.
Ref.~\cite{Kourkoulou:2017zaj} introduces the flip group,
a subgroup of \(O(N)\) generated by \(N/2\) transformations,
one for each bilinear \(S_k\).
The generator associated with \(S_k\) sends
\(\psi_{2k}\to-\psi_{2k}\) and \(j_I\to-j_I\)
for \(2k\in I\), while leaving \(\psi_l\) unchanged
for \(l\neq2k\) and \(j_I\) unchanged for \(2k\notin I\).
These transformations map one choice of signs \(s\)
to another, \(s'\), and hence map \(|B_s\rangle\)
to \(|B_{s'}\rangle\).
For arbitrary \(i\) and \(\ell\), flavour-diagonal products \(\psi_i(t_1)\psi_i(t_2)\) and combinations 
\(\psi_{2\ell-1}(t_1)\psi_{2\ell}(t_2)S_\ell(0)\)
are invariant under every generator of the flip group.
This invariance can be used to relate the expectation values
of these operators in different KM states.
For either flip-invariant operator \(\mathcal O\), a flip
gives, before disorder averaging,
\begin{align}
    \langle B_s|\mathcal O[j]|B_s\rangle
    &=\langle B_{s'}|\mathcal O[j']|B_{s'}\rangle,
\end{align}
where \(j'=\{j'_I\}\) denotes the set of couplings
after the flip.
Since the coupling distribution \eqref{eq:SM-covariance}
is invariant under these sign reversals, replacing \(j'\)
by \(j\) leaves the disorder average unchanged.
Thus,
\begin{align}
    \overline{\langle B_s|\mathcal O[j]|B_s\rangle}
    &=\overline{\langle B_{s'}|\mathcal O[j]|B_{s'}\rangle}.
\end{align}
Since the flip group relates all choices of signs,
the disorder-averaged expectation value in any KM state
\(|B_s\rangle\) equals its average over the KM basis.
Using also Eq.~\eqref{eq:KM-completeness}, we obtain
\begin{align}
    \overline{\langle B_s|\mathcal O[j]|B_s\rangle}
    &=2^{-N/2}\sum_{s'}
    \overline{\langle B_{s'}|\mathcal O[j]|B_{s'}\rangle}
    \nonumber\\
    &=\overline{\langle\mathcal O[j]\rangle_{\beta=0}}.
\label{eq:KM-flip-invariant-expectation}
\end{align}
Thus, for every flip-invariant operator \(\mathcal O\),
the disorder-averaged expectation value in any KM state
\(|B_s\rangle\) coincides with its disorder-averaged
thermal equilibrium expectation value at \(\beta=0\).

Applying this identity to the flavour-diagonal products
\(\psi_i(t_1)\psi_i(t_2)\) gives
\begin{align}
    \overline{\langle B_s|
    \psi_i(t_1)\psi_i(t_2)|B_s\rangle}
    &=\imath G^{>}(t_1,t_2)+\mathcal O(N^{-1}),
\end{align}
where \(G^{>}=-\frac{\imath}{2}e^{g/q}\) is the thermal
Green function obtained from
Eq.~\eqref{eq:diagonal_solution_main} at \(\beta=1/T_0=0\),
corresponding to \(\theta_0=0\).
Applying the same identity to \(S_k(t)S_k(0)\) and using
\(S_k(0)|B_s\rangle=s_k|B_s\rangle\), we obtain
\begin{align}
    s_k\overline{\langle B_s|S_k(t)|B_s\rangle}
    &=\overline{\langle S_k(t)S_k(0)\rangle_{\beta=0}}
    \nonumber\\
    &=\left[2\imath G^{>}(t,0)\right]^2
    +\mathcal O(N^{-1}),
    \label{eq:KM-bilinear-propagation}
\end{align}
where the last equality uses the standard large-\(N\)
factorization of thermal SYK \(n\)-point functions
into products of two-point functions \cite{Gross:2017aos}.

To determine which flavour-off-diagonal two-point functions
can be nonzero, consider a simultaneous sign reversal
of both fermions in a single pair $(2k-1,2k)$.
This sends
\(\psi_{2k-1}\to-\psi_{2k-1}\) and
\(\psi_{2k}\to-\psi_{2k}\), leaving the other fermions
unchanged. The accompanying coupling transformation
reverses the sign of \(j_I\) whenever \(I\) contains
exactly one of these two indices, leaving the other
couplings unchanged.  
Every \(S_\ell\) remains invariant, so \(|B_s\rangle\)
is preserved up to a phase.
Consider the product \(\psi_i(t_1)\psi_j(t_2)\), where
the flavour indices \(i\) and \(j\) belong to different
pairs of the form \((2k-1,2k)\).
Applying this transformation to the pair containing \(i\)
reverses the sign of this product.
Since the coupling distribution
\eqref{eq:SM-covariance} is invariant under these
sign reversals, the disorder-averaged correlator
must vanish:
\begin{align}
    \overline{\langle B_s|
    \psi_i(t_1)\psi_j(t_2)|B_s\rangle}
    =-\overline{\langle B_s|
    \psi_i(t_1)\psi_j(t_2)|B_s\rangle}
    =0,
\end{align}
whenever \(i\) and \(j\) belong to different pairs.

Invoking self-averaging, the disorder-averaged correlators
coincide with those in a typical realization of the couplings
at leading large \(N\) \cite{Kourkoulou:2017zaj}.
Using also \(s_k^2=1\), we obtain
\begin{align}
    \langle B_s|\psi_i(t_1)\psi_i(t_2)|B_s\rangle
    &=\imath G^{>}(t_1,t_2)+\mathcal O(N^{-1}),
    \label{eq:KM-diagonal-correlator}\\
    \langle B_s|\psi_i(t_1)\psi_j(t_2)|B_s\rangle
    &=\mathcal O(N^{-1}),
    \quad i,j\text{ in different pairs},
    \label{eq:KM-interpair-correlator}\\
    \langle B_s|S_k(t)|B_s\rangle
    &=s_k\left[2\imath G^{>}(t,0)\right]^2
    +\mathcal O(N^{-1}). \label{skt}
\end{align}
The same flip-group argument, together with the usual
large-\(N\) factorization in SYK, extends these results
to higher-point correlation functions.
The KM states therefore reproduce the correlation
functions of thermal SYK at leading large \(N\), with
one class of exceptions:
the correlators sensitive to the initial pair
polarizations \(S_k\) or their products.
In particular, the polarization
\(\langle B_s|S_k(\delta_t)|B_s\rangle\)
provides an NEQ witness analogous to that in
Eq.~\eqref{eq:probe} of the main text.
As seen from Eq.~\eqref{skt}, this NEQ witness decays as
$[2\imath G^{>}(\delta_t,0)]^2$, where $\delta_t$
is the time elapsed since preparation.
This is reminiscent of the return-quench NEQ witness
in the main text.
As we show below, the $q$-body analogue instead decays
as $[2\imath G^{>}(\delta_t,0)]^q$, reproducing the
decay of the NEQ witness in
Eq.~\eqref{eq:exact_return_result} for a quench between
uncorrelated Hamiltonians, $c_{01}=0$.

\subsection{Comparison with the Quenched State}
\label{sec:qbody-KM}
We now construct a \(q\)-body analogue of the KM state
and use it as the initial state for evolution under the 
SYK Hamiltonian \(H_{\mathrm{SYK}}\).
This provides an NEQ witness that is a \(q\)-body
analogue of the pair polarizations \(S_k\)
in Eq.~\eqref{skt}.
We keep only the leading terms in \(1/q\) and \(q^2/N\), unless stated otherwise. 

For simplicity, we assume that \(N\) is divisible by \(q\) and partition the indices \(k=1,\ldots,N/2\) into sets \(A_1,\ldots,A_{N/q}\), each containing \(q/2\) indices.
Define
\begin{align}
    Q_{A_\alpha}&=\prod_{k\in A_\alpha} S_k=    (- 2\imath)^{q/2}    \prod_{k\in A_\alpha}    \psi_{2k-1}\psi_{2k}.
\label{eq:qbody-KM-polarization}
\end{align}
They satisfy
\begin{align}
    Q_{A_\alpha}^2&=1,\quad[Q_{A_\alpha},Q_{A_\beta}]=0,
\end{align}
and define the preparation Hamiltonian
\begin{align}
    H_{\mathrm{pr}}&=-\sum_{\alpha=1}^{N/q}  K_\alpha Q_{A_\alpha} ,
    \label{eq:qbody-KM-parent}
\end{align}
with real couplings \(K_\alpha\).
Its thermal state is
\begin{align}
    \rho_{\mathrm{pr}}
    &=
    \frac{
        e^{-\beta_{\mathrm{pr}}H_{\mathrm{pr}}}
    }{
        \Tr e^{-\beta_{\mathrm{pr}}H_{\mathrm{pr}}}
    }
    \label{eq:qbody-KM-state}
\end{align}
Defining
\begin{align}
    Q_{A_\alpha}|B_s\rangle
    &=\sigma_\alpha(s)|B_s\rangle,\quad\sigma_\alpha(s)
    =\prod_{k\in A_\alpha}s_k,\\
    p_\alpha
    &=
    \tanh\left(
        \beta_{\mathrm{pr}}K_\alpha
    \right),
    \\
    w_s
    &=
    2^{-N/2}
    \prod_{\alpha=1}^{N/q}
    [1+p_\alpha \sigma_\alpha(s)], 
\end{align}
we can represent \(\rho_\mathrm{pr}\) as 
\begin{align}
    \rho_{\mathrm{pr}}
    &=
    2^{-N/2}
    \prod_{\alpha=1}^{N/q}
    \left(
        1+p_\alpha Q_{A_\alpha}
    \right)
    \nonumber\\
    &=
    \sum_s
    w_s
    |B_s\rangle\langle B_s|.
\label{eq:qbody-KM-decomposition}
\end{align}
For \(q>2\), the state \(\rho_{\mathrm{pr}}\) is a correlated mixture of KM states.\footnote{For \(q=2\) and \(\beta_\mathrm{pr}\to \infty\) this construction yields the unfiltered KM state \(|B_s\rangle\). }
The \(q\)-body operators \(Q_{A_\alpha}\), rather than
the individual bilinears \(S_k\), carry the polarization.
We choose the initial state \(\rho(0)=\rho_{\mathrm{pr}}\) and evolve with a \(q\)-body SYK Hamiltonian until \(t=\delta_t\).
The protocol reads
\begin{align}
    H(t)
    &=
    \begin{cases}
        H_{\mathrm{pr}},
        &t<0,
        \\[1mm]
         H_{\mathrm{SYK}},
        &0<t<\delta_t,
        \\[1mm]
        H_{\mathrm{pr}},
        &t>\delta_t,
    \end{cases}
    \label{eq:qbody-KM-protocol}
\end{align}
where the return quench plays the role of reading out the surviving memory of the initial state \(\rho(0)=\rho_{\mathrm{pr}}\).

The Hamiltonian \(H_{\mathrm{SYK}}\) is statistically independent
of \(H_{\mathrm{pr}}\), and therefore of the weights
\(w_s\).
The flip-group arguments in
Eqs.~\eqref{eq:KM-completeness}--\eqref{eq:KM-interpair-correlator}
apply term by term to expectation values in the
decomposition \eqref{eq:qbody-KM-decomposition}
of \(\rho_{\mathrm{pr}}\).
Since \(\sum_s w_s=1\), the correlation functions
evaluated in \(\rho_{\mathrm{pr}}\) under evolution
with \(H_{\mathrm{SYK}}\) coincide with those
evaluated in the maximally mixed state, with the
same class of exceptions as for the KM states:
the correlators sensitive to the initial pair
polarizations \(S_k\).

The mixture \eqref{eq:qbody-KM-decomposition}
further restricts which polarizations can be nonzero.
For \(P_B(t)=\prod_{k\in B}S_k(t)\), the operator
\(P_B(t)P_B(0)\) is flip invariant.
Since \(P_B(0)\) has eigenvalue
\(\prod_{k\in B}s_k\) in \(|B_s\rangle\),
the flip-group argument
\eqref{eq:KM-flip-invariant-expectation} gives
\begin{align}
    \overline{\langle P_B(t)\rangle}
    &=
    \left(\sum_s w_s\prod_{k\in B}s_k\right)
    \overline{\langle P_B(t)P_B(0)\rangle_{\beta=0}},
\label{eq:qbody-KM-polarization-propagation}
\end{align}
for \(0\le t\le\delta_t\).

Suppose \(B\) contains some but not all indices
of a block \(A_\alpha\) from
\eqref{eq:qbody-KM-polarization}.
Choose \(\ell\in B\cap A_\alpha\) and
\(k\in A_\alpha\setminus B\).
Reversing \(s_\ell\) and \(s_k\) leaves every
\(\sigma_\alpha(s)\), and hence \(w_s\), unchanged,
while it reverses the polarization eigenvalue
\(\prod_{r\in B}s_r\).
Since \(s\) is a dummy summation index in
\eqref{eq:qbody-KM-polarization-propagation},
the sum vanishes for these products \(P_B\).
Thus, \(\overline{\langle P_B(t)\rangle}\) can be
nonzero only if \(P_B(t)\) is a product of the
\(Q_{A_\alpha}(t)\).

For an operator $A$, we write
$\langle A\rangle(t)\equiv\Tr[A\rho(t)]$.
In the initial state $\rho(0)=\rho_{\mathrm{pr}}$,
the expectation values of the polarization operators
$Q_{A_\alpha}$ and the Hamiltonian $H_{\mathrm{pr}}$ are
\begin{align}
    \langle Q_{A_\alpha}(0)\rangle
    &=p_\alpha,
    \quad
    \langle H_{\mathrm{pr}}\rangle(0)
    =-\sum_{\alpha=1}^{N/q}K_\alpha p_\alpha.
    \label{eq:qbody-KM-initial-energy}
\end{align}

Next, we invoke self-averaging and use the same
large-\(N\) factorization as in
\eqref{eq:KM-bilinear-propagation}.
Omitting subleading terms as in the main text,
we obtain from \eqref{eq:qbody-KM-polarization-propagation}
\begin{align}
    \langle Q_{A_\alpha}(t)\rangle
    &=p_\alpha\left[2\imath G^{>}(t,0)\right]^q
    \nonumber\\
    &=p_\alpha e^{g(t,0)}
    \nonumber\\
    &=\langle Q_{A_\alpha}(0)\rangle e^{g(t,0)},
    \label{eq:qbody-KM-factorization}
\end{align}
for \(0\le t\le\delta_t\), where \(g\) is the
\(\beta=1/T_0=0\) solution \eqref{eq:diagonal_solution_main}.
As established by the flip-group argument above,
the correlation functions of \(\rho(t)\) coincide
with those evaluated in the maximally mixed state,
except for correlators sensitive to the initial
\(Q_{A_\alpha}\) polarizations.

The return quench at \(t=\delta_t\) therefore reveals
the surviving polarization through its contribution
to the energy after the return:
\begin{align}
    \langle H_{\mathrm{pr}}\rangle(\delta_t)
    &=
    -\sum_{\alpha=1}^{N/q}
    K_\alpha
    \langle Q_{A_\alpha}(\delta_t)\rangle
    \, \nonumber\\
    &=\langle H_{\mathrm{pr}}\rangle(0)
    e^{g(\delta_t,0)}.
\end{align}
After returning to \(H_{\mathrm{pr}}\), every \(Q_{A_\alpha}\)
is conserved because \([Q_{A_\alpha},H_{\mathrm{pr}}]=0\). 
The return therefore does not recreate the correlations:
it makes the surviving polarization contribute to the
energy and then freezes it.

We now compare this result with the return protocol
of the main text.
To match the statistical independence of
\(H_{\mathrm{SYK}}\) and the initial state
\(\rho_{\mathrm{pr}}\), we set
\(c_{01}=c_{12}=0\).
This gives \(\beta_1=1/T_1=0\), corresponding to
\(\theta_1=0\). 
This does not mean that the quenched state is
maximally mixed.
Rather, \(\beta_1\) is the inverse temperature of
the thermal reference state whose correlation
functions coincide with the post-quench correlators
\eqref{eq:diagonal_solution_main}.
At \(\beta_1=0\), this reference state is maximally
mixed, yet the return quench distinguishes it from
the quenched state for \(\beta_0>0\).
The exact return result \eqref{eq:exact_return_result}
then reduces to
\begin{align}
    \langle H_0\rangle(\delta_t)
    = & e^{g_{11}(\delta_t,0)} \langle H_0\rangle(0).
    \label{eq:qbody-KM-ordinary-return}
\end{align}
Thus, the generalized KM return protocol and the
ordinary SYK return exhibit the same \(q\)-leg
decay \eqref{eq:qbody-KM-factorization}.

The difference lies in how the initial polarization is distributed.
The preparation Hamiltonian \(H_{\mathrm{pr}}\)
contains \(N/q\) selected operators \(Q_{A_\alpha}\),
each with an order-one initial expectation value
\(p_\alpha\).
By contrast, writing
\begin{align}
    H_0
    &=
    \imath^{q/2}
    \sum_{I=1}^{\binom{N}{q}}
    j_{0I}\psi_I,
    &
    \psi_I
    &=
    \psi_{i_1}\cdots\psi_{i_q},
\end{align}
the polarization relevant to the return is distributed
over \(\binom{N}{q}\) statistically equivalent
\(q\)-body operators \(\psi_I\).
Since the return energy is extensive, the
contribution of each operator \(\psi_I\) scales as
\begin{align}
    j_{0I}\langle\psi_I\rangle
    &\sim
    \frac{N}{\binom{N}{q}}
    \sim
    N^{-(q-1)}.
\end{align}
Since \(j_{0I}\sim N^{-(q-1)/2}\), this
energy contribution is carried by a polarization 
\(\langle\psi_I\rangle\) of order \(N^{-(q-1)/2}\).

In the KM states and their \(q\)-body analogue,
the memory is visible in individual nonthermal
correlators \eqref{eq:KM-bilinear-propagation}
and~\eqref{eq:qbody-KM-factorization}
at leading large \(N\).
In the post-quench state of the main text, the memory
resides in \(\binom{N}{q}\) correlators
\(\langle\psi_I\rangle\), which are already suppressed
in the initial equilibrium state of ordinary SYK.
Their surviving collective contribution nevertheless
produces an extensive return energy.
The usual fermion correlation functions are thus
exactly thermal at leading large \(N\), yet the
state retains extensive nonthermal memory.

\section{Finite-q and the non-thermal correlations}
\label{Finite-q and the non-thermal correlations}
We now revisit the return-quench protocol
$H_0\to H_1\to H_0$ of the main text
(Fig.~\ref{fig:protocol}).
We first show that the single-quench energy relation
\eqref{single_quench_result} holds for arbitrary
even $q$ in the large-$N$ limit.
Next, we specialize to the regime of weak quenches,
$1-c_{01}^2\ll1$.
In this regime, we can explicitly see how
the nonthermal contributions to $\langle\psi_I\rangle$
are individually suppressed by inverse powers of $N$
and how their sum over all $\binom{N}{q}$ suppressed
terms gives rise to the extensive NEQ witness
\eqref{eq:probe}.

\subsection{Finite-$q$ energy relation}
We start from the Gibbs state of $H_0$ at inverse
temperature $\beta_0=1/T_0$.
The first quench occurs at $t=\tau_1=0$, and
the return quench at $t=\tau_2=\delta_t$.
We denote Gibbs states and expectation values
in these states by
\begin{equation}
 \rho_{\beta,H}\equiv\frac{e^{-\beta H}}{Z_\beta[H]},
 \qquad
 \langle A\rangle_{\beta,H}
 \equiv\Tr[\rho_{\beta,H}A].
\end{equation}

Since the coupling vectors are chosen to have equal norms
as in the main text,
$|\vec{\mathcal J}_0|=|\vec{\mathcal J}_1|=\mathcal J$,
Eq.~\eqref{eq:SM-covariance} gives
$\overline{j_{0I}j_{1J}}
=c_{01}\overline{j_{1I}j_{1J}}$.
Defining
\begin{equation}
 j_{\perp I}\equiv j_{0I}-c_{01}j_{1I},
\end{equation}
we obtain
\begin{equation}
 \overline{j_{\perp I}j_{1J}}
 =
 \overline{j_{0I}j_{1J}}
 -c_{01}\overline{j_{1I}j_{1J}}
 =0.
\label{eq:SM-perp-cross-covariance}
\end{equation}
Since $j_{\perp I}$ is a linear combination of the
original Gaussian couplings $j_{0I}$ and $j_{1I}$,
its joint distribution with $j_{1I}$ is Gaussian.
Their vanishing cross-covariance
\eqref{eq:SM-perp-cross-covariance} therefore implies
that the couplings $j_{\perp I}$ are
independent of those of $H_1$.

The definition of $j_{\perp I}$ gives the decomposition
\begin{equation}
 H_0=c_{01}H_1+H_\perp,
 \qquad
 H_\perp\equiv
 \imath^{q/2}\sum_I j_{\perp I}\psi_I,
 \label{eq:SM-H-decomposition}
\end{equation}
The first quench of the return protocol
(Fig.~\ref{fig:protocol}), at $t=\tau_1=0$,
thus removes $H_\perp$ and rescales the $H_1$ component.
The second quench, at $t=\tau_2=\delta_t$, restores
$H_\perp$ and reverses this rescaling.
The rescalings in the protocol are dynamically
trivial: an overall rescaling of a Hamiltonian
only changes the clock of its unitary evolution,
$e^{-\imath aH_1t}=e^{-\imath H_1(at)}$ \cite{Kourkoulou:2017zaj}. 

The independent couplings $j_{\perp I}$ have covariance
\begin{equation}
 \overline{j_{\perp I}j_{\perp J}}
 =(1-c_{01}^2)\overline{j_{1I}j_{1J}}.
 \label{eq:SM-perp-covariance}
\end{equation}
corresponding to
\begin{equation}
 |\vec{\mathcal J}_\perp|^2
 =(1-c_{01}^2)\mathcal J^2,
 \qquad
 \vec{\mathcal J}_\perp
 =\vec{\mathcal J}_0-c_{01}\vec{\mathcal J}_1.
\label{eq:SM-perp-strength}
\end{equation}

Invoking the usual self-averaging arguments
\cite{Gross:2017aos}, the expectation value
$\langle H_1\rangle_{\beta_0,H_0}$ in a typical
realization of the couplings $j_{0I}$ and $j_{1I}$
agrees with its disorder average at large $N$.
We therefore derive the finite-$q$ single-quench
energy relation \eqref{single_quench_result}
using disorder-averaged expectation values.
Interchanging the roles of $H_0$ and $H_1$ in
Eq.~\eqref{eq:SM-H-decomposition}, we write
\begin{equation}
 H_1=(H_1-c_{01}H_0)+c_{01}H_0,
\end{equation}
where the Hamiltonian $H_1-c_{01}H_0$ has couplings
$j_{1I}-c_{01}j_{0I}$.
Using
$\overline{j_{1I}j_{0J}}
=c_{01}\overline{j_{0I}j_{0J}}$
from Eq.~\eqref{eq:SM-covariance}, we find
\begin{align}
 \overline{(j_{1I}-c_{01}j_{0I})j_{0J}}
 &=
 \overline{j_{1I}j_{0J}}
 -c_{01}\overline{j_{0I}j_{0J}}
 =0.
\end{align}
Thus, the couplings $j_{1I}-c_{01}j_{0I}$ are
independent of those of $H_0$ and have zero mean
by linearity.
Since the Gibbs state $\rho_{\beta_0,H_0}$
depends only on $H_0$, we can average
$H_1-c_{01}H_0$ while holding $H_0$ and
therefore $\rho_{\beta_0,H_0}$ fixed.
This contribution then vanishes, giving
\begin{align}
 \overline{\langle H_1\rangle_{\beta_0,H_0}}
 ={}&
 c_{01}\,
 \overline{\langle H_0\rangle_{\beta_0,H_0}}
 +
 \overline{\langle H_1-c_{01}H_0
 \rangle_{\beta_0,H_0}}
 \notag\\
 ={}&
 c_{01}\,
 \overline{\langle H_0\rangle_{\beta_0,H_0}}.
\label{eq:SM-single-quench-finite-q}
\end{align}
With the normalization
\eqref{eq:SM-energy-finite-q},
$\epsilon_a=2q^2\langle H_a\rangle_{\beta_0,H_0}/N$,
the result reads $\epsilon_1=c_{01}\epsilon_0$.
This reproduces the single-quench energy relation
\eqref{single_quench_result} for arbitrary even $q$
in the large-$N$ limit.

\subsection{Nonthermal correlations and weak quenches}
\label{sec:SM-nonthermal-correlations}
We now specialize to weak quenches,
$s_{01}\equiv\sqrt{1-c_{01}^2}\ll1$,
with the initial Gibbs state $\rho_{\beta_0,H_0}$
at an arbitrary temperature $T_0>0$.
We keep $q$ finite until we explicitly invoke
the large-$q$ limit below. 
We begin by expanding the initial Gibbs state
$\rho_{\beta_0,H_0}$ about $\rho_{\beta_0,H_1}$
in powers of $s_{01}$.
Since $c_{01}-1=\mathcal O(s_{01}^2)$,
the rescaling of $H_1$ contributes only at second order.
Since $H_\perp=\mathcal O(s_{01})$, the first-order
expansion reads
\begin{align}
 \rho_{\beta_0,H_0}
 ={}&\rho_{\beta_0,H_1}
 -\beta_0\int_0^1du\,
 (\rho_{\beta_0,H_1})^{1-u}
 \notag\\
 &\quad\times
 \bigl[H_\perp-\langle H_\perp\rangle_{\beta_0,H_1}\bigr]
 (\rho_{\beta_0,H_1})^u
 +\mathcal O(s_{01}^2).
 \label{eq:SM-weak-state-expansion}
\end{align}
where \(\langle H_\perp\rangle_{\beta_0,H_1}\)
arises from expanding the partition function
\(Z_{\beta_0}[H_0]\) about \(Z_{\beta_0}[H_1]\);
see the derivation leading to Eq.~(35)
in Ref.~\cite{Kliesch2014}.

Using \(H_1\) to generate the time evolution, we compare
the expectation value of \(\imath^{q/2}\psi_I\)
at \(t=\delta_t\), starting from the initial state
\(\rho_{\beta_0,H_0}\), with its value in the thermal
reference state \(\rho_{\beta_0,H_1}\).
The thermal reference state is stationary
under $H_1$, so
$\langle\psi_I(\delta_t)\rangle_{\beta_0,H_1}
=\langle\psi_I\rangle_{\beta_0,H_1}$.
Setting \(\tau=\beta_0u\) in
Eq.~\eqref{eq:SM-weak-state-expansion}, we obtain:
\begin{align}
 &\imath^{q/2}\bigl[
 \langle\psi_I(\delta_t)\rangle_{\beta_0,H_0}
 -\langle\psi_I\rangle_{\beta_0,H_1}
 \bigr]
 \label{eq:SM-weak-polarization}\\
 &\quad=-\imath^q\sum_J j_{\perp J}
 \int_0^{\beta_0}d\tau\,
 \langle\psi_J(-\imath\tau)\psi_I(\delta_t)
 \rangle_{\beta_0,H_1}
 +\cdots .\notag
\end{align}
The ellipses denote subleading corrections in
$s_{01}$ and $1/N$, and the sum over multi-indices $J$
follows from inserting the expression for $H_\perp$
in Eq.~\eqref{eq:SM-H-decomposition}.

The independence of $j_{1I}$ from $j_{\perp I}$
allows the disorder average over $j_{1I}$ in
Eq.~\eqref{eq:SM-weak-polarization} to be
performed at fixed $j_{\perp I}$.
Using the standard large-$N$ SYK factorization
into $q$ fermion propagators
\cite{Gross:2017aos}, we obtain
\begin{align}
 &\imath^q
 \langle\psi_J(-\imath\tau)\psi_I(\delta_t)
 \rangle_{\beta_0,H_1}
 \notag\\
 &\quad=
 \frac{\delta_{IJ}}{2^q}
 \bigl[2\imath G_{\beta_0,H_1}^{>}
 (-\imath\tau,\delta_t)\bigr]^q +\cdots .
\label{thermal-expectation-value-of-psis}
\end{align}
Here $G_{\beta_0,H_1}^{>}$ is the fermion Green
function in the Gibbs state $\rho_{\beta_0,H_1}$,
with time evolution generated by $H_1$.
At large $q$, it is determined by the analytic continuation
of the main-text result \eqref{eq:diagonal_solution_main}, while for general finite $q$ it is not known in closed form.

Inserting Eq.~\eqref{thermal-expectation-value-of-psis}
into Eq.~\eqref{eq:SM-weak-polarization}
and using $\delta_{IJ}$ to collapse the sum over
multi-indices $J$, we obtain
\begin{align}
&\imath^{q/2}\bigl[
 \langle\psi_I(\delta_t)\rangle_{\beta_0,H_0}
 -\langle\psi_I\rangle_{\beta_0,H_1}
 \bigr]
 \notag\\
 &\quad=-\frac{j_{\perp I}}{2^q}
 \int_0^{\beta_0}d\tau\,
 \bigl[2\imath G_{\beta_0,H_1}^{>}
 (-\imath\tau,\delta_t)\bigr]^q
 +\cdots .
 \label{eq:SM-weak-integral}
\end{align}
The thermal one-point function
$\langle\psi_I\rangle_{\beta_0,H_1}$ is of order
$N^{-(q-1)/2}$, while the displayed first-order
deviation is of order $s_{01}N^{-(q-1)/2}$.
Although both are suppressed, the contributions
of these deviations to the NEQ witness \eqref{eq:probe}
add up to an extensive quantity, as we now show.

Taking the expectation value of the decomposition
\eqref{eq:SM-H-decomposition} in the quenched state
$\rho(\delta_t)=e^{-\imath H_1\delta_t}
\rho_{\beta_0,H_0}e^{\imath H_1\delta_t}$ gives
\begin{equation}
 \langle H_0\rangle(\delta_t)
 -c_{01}\langle H_1\rangle(\delta_t)
 =\langle H_\perp\rangle(\delta_t).
 \label{eq:SM-perp-witness}
\end{equation}
Since the couplings $j_{\perp I}$ have zero mean
and, by Eq.~\eqref{eq:SM-perp-cross-covariance},
are independent of those of $H_1$, the expectation value 
$\langle H_\perp\rangle_{\beta,H_1}$ vanishes
upon disorder averaging for any $\beta$.
A nonzero right-hand side of Eq.~\eqref{eq:SM-perp-witness}
therefore provides a NEQ witness.
The post-quench energy $\langle H_1\rangle(\delta_t)$
is conserved and given by
Eq.~\eqref{eq:SM-single-quench-finite-q}.
Thus, the waiting-time dependence of
$\langle H_0\rangle(\delta_t)$ enters entirely through
$\langle H_\perp\rangle(\delta_t)$.

Multiplying Eq.~\eqref{eq:SM-weak-integral} by
$j_{\perp I}$ and summing over $I$ gives
\begin{align}
 &\imath^{q/2}\sum_I j_{\perp I}
 \bigl[
 \langle\psi_I(\delta_t)\rangle_{\beta_0,H_0}
 -\langle\psi_I\rangle_{\beta_0,H_1}
 \bigr]
\notag \\
 &\quad=
 \langle H_\perp\rangle(\delta_t)
 -\langle H_\perp\rangle_{\beta_0,H_1}.
\label{eq:SM-summed-polarization}
\end{align}
Again, $\langle H_\perp\rangle_{\beta_0,H_1}$
vanishes upon disorder averaging.
We combine Eqs.~\eqref{eq:SM-weak-integral},
\eqref{eq:SM-perp-witness},
and \eqref{eq:SM-summed-polarization}.
Invoking self-averaging and using
the covariance and coupling strength in
Eqs.~\eqref{eq:SM-perp-covariance}
and \eqref{eq:SM-perp-strength}, together with the
coupling normalization in Eq.~\eqref{eq:SM-covariance},
we obtain
\begin{align}
 \langle H_0&\rangle(\delta_t)
 -c_{01}\langle H_1\rangle(\delta_t)
 \label{eq:SM-weak-return-finite-q}
\\
 &=-\frac{Ns_{01}^2\mathcal J^2}{2q^2}
 \int_0^{\beta_0}d\tau\,
 \bigl[2\imath G_{\beta_0,H_1}^{>}
 (-\imath\tau,\delta_t)\bigr]^q
 +\cdots ,
\notag
\end{align}
where the ellipses denote subleading corrections in
$s_{01}$ and $1/N$.
Using the energy normalization
\eqref{eq:SM-energy-finite-q},
$\epsilon_2(\delta_t)=2q^2\langle H_0\rangle(\delta_t)/N$
and $\epsilon_1=2q^2\langle H_1\rangle(\delta_t)/N$,
Eq.~\eqref{eq:SM-weak-return-finite-q} becomes
\begin{equation}
 \boxed{
 \begin{aligned}
 &\epsilon_2(\delta_t)-c_{01}\epsilon_1\\
 &\quad=-s_{01}^2\mathcal J^2
 \int_0^{\beta_0}d\tau\,
 \bigl[2\imath G_{\beta_0,H_1}^{>}
 (-\imath\tau,\delta_t)\bigr]^q
 +\cdots .
 \end{aligned}
 }
 \label{eq:SM-weak-return-epsilon}
\end{equation}
In the weak-quench limit,
Eq.~\eqref{eq:SM-weak-return-epsilon} determines
the NEQ witness from the thermal two-point function
$G_{\beta_0,H_1}^{>}$.
The formula is valid for any even $q$ in the
large-$N$ limit considered here.
For finite $q$ at finite temperature,
$G_{\beta_0,H_1}^{>}$ generally has to be obtained
numerically.
In the large-$q$ limit, it is known in closed form,
as given in Eq.~\eqref{eq:diagonal_solution_main}.
Analytically continuing its first time argument
to $-\imath\tau$, we obtain
\begin{align}
 &\bigl[2\imath G_{\beta_0,H_1}^{>}
 (-\imath\tau,\delta_t)\bigr]^q
 \notag\\
 &\quad=
 \left[
 \frac{2\theta_0T_0/\mathcal J}
 {\cosh\!\left[
 2\theta_0T_0(-\delta_t-\imath\tau)
 +\imath\theta_0
 \right]}
 \right]^2
 +\mathcal O(q^{-1}).
 \label{eq:SM-weak-thermal-propagator}
\end{align}
Since $H_0$ and $H_1$ have the same coupling strength
$\mathcal J$, $G_{\beta_0,H_1}^{>}$ is parametrized
by the initial thermal parameters $\theta_0$ and $T_0$.
At leading order in $1/q$, inserting
Eq.~\eqref{eq:SM-weak-thermal-propagator}
into Eq.~\eqref{eq:SM-weak-integral}
and performing the imaginary-time integral yields
\begin{align}
 &\imath^{q/2}\bigl[
 \langle\psi_I(\delta_t)\rangle_{\beta_0,H_0}
 -\langle\psi_I\rangle_{\beta_0,H_1}
 \bigr]
 \notag\\
 &\quad=-\frac{j_{\perp I}}{2^{q-1}}
 \frac{\sin\theta_0}{\mathcal J}
 \frac{\cos^2\theta_0}
 {\cos^2\theta_0+
 \sinh^2(2\theta_0T_0\delta_t)}
 +\cdots .
 \label{eq:SM-weak-integral-large-q}
\end{align}
The large-$q$ form of
Eq.~\eqref{eq:SM-weak-return-epsilon} thus reads
\begin{align}
 \epsilon_2(\delta_t)-c_{01}\epsilon_1
 ={}&s_{01}^2\,\epsilon_0
 \frac{\cos^2\theta_0}
 {\cos^2\theta_0+
 \sinh^2(2\theta_0T_0\delta_t)}+\cdots ,
 \label{eq:SM-weak-return}
\end{align}
with the ellipses denoting subleading corrections in
$s_{01}$, $1/q$, and $1/N$.

Since $c_{12}=c_{01}$ and
$\epsilon_1=c_{01}\epsilon_0$ by
Eq.~\eqref{eq:SM-single-quench-finite-q},
we have
$\epsilon_0-c_{12}\epsilon_1=s_{01}^2\epsilon_0$.
At fixed $T_0>0$, the thermal parameters
$(\theta_1,T_1)$ differ from $(\theta_0,T_0)$
only at order $s_{01}^2$.
Replacing them in Eq.~\eqref{eq:SM-weak-return}
therefore changes the witness only at order
$s_{01}^4$, allowing us to rewrite it as
\begin{align}
 \epsilon_2(\delta_t)
 ={}&c_{12}\epsilon_1
 +[\epsilon_0-c_{12}\epsilon_1]
 \left|e^{g_{11}(\delta_t,0)}\right|
 +\cdots .
\end{align}
This reproduces the main-text return result
\eqref{eq:exact_return_result}, evaluated at
$\tau_1=0$ and $\tau_2=\delta_t$,
to the order considered.

Equations~\eqref{eq:SM-weak-integral} and
\eqref{eq:SM-weak-integral-large-q} show explicitly
how the deviations from the thermal reference
in the individual $q$-body correlations decay.
The thermal expectation values
$\langle\psi_I\rangle_{\beta_0,H_1}$ are already
of order $N^{-(q-1)/2}$, while the nonthermal
deviations are of order $s_{01}N^{-(q-1)/2}$.
The usual leading large-$N$ treatment of individual
fermion correlators is therefore blind to these
$q$-body expectation values and hence also to
their deviations from the thermal reference.
The contributions of these deviations nevertheless
add up to an extensive quantity.

\end{document}